\documentclass[nohyper]{JHEP3}

\usepackage{cite}
\usepackage{epsfig}

\newcommand{\mycomm}[1]{\hfill\break $\phantom{a}$\kern-3.5em{\tt===$>$ \bf
#1}\hfill\break}
\newcommand{\mycommA}[1]{\hfill\break $\phantom{a}$\kern-3.5em{\tt   $>$ \bf
#1}\hfill\break}

\newcommand{\be}{\begin{equation}}
\newcommand{\ee}{\end{equation}}
\newcommand{\ba}{\begin{eqnarray}}
\newcommand{\ea}{\end{eqnarray}}

\def\eq#1{Eq.~(\ref{#1})}

\def\MSbar{\hbox{\tiny ${\overline{\rm MS}}$}}

\def\tot{\hbox{\tiny total}}

\def\cusp{\hbox{\tiny cusp}}
\def\eff{\hbox{\tiny eff}}
\def\PT{\hbox{\tiny PT}}
\def\PV{\hbox{\tiny PV}}

\def\NP{\hbox{\tiny NP}}

\catcode`\@=11 
\def\lsim{\mathrel{\mathpalette\@versim<}}
\def\gsim{\mathrel{\mathpalette\@versim>}}
\def\@versim#1#2{\vcenter{\offinterlineskip
        \ialign{$\m@th#1\hfil##\hfil$\crcr#2\crcr\sim\crcr } }}
\catcode`\@=12 

\newcommand{\ysl}{\mbox{$y$\hspace{-0.5em}\raisebox{0.1ex}{$/$}}}
\newcommand{\vsl}{\mbox{$v$\hspace{-0.5em}\raisebox{0.1ex}{$/$}}}

\title{Radiative and semi-leptonic B--meson decay spectra:
Sudakov resummation beyond logarithmic accuracy and the pole mass}

\author{Einan Gardi \\
Cavendish Laboratory, University of Cambridge\\
Madingley Road, Cambridge, CB3 0HE, UK}

\abstract{The inclusive spectra of radiative and semi-leptonic
B--meson decays near the endpoint is computed taking into account
renormalons in the Sudakov exponent (Dressed Gluon Exponentiation).
In this framework we demonstrate the factorization of decay spectra
into hard, jet and soft functions and discuss the universality of
the latter two. Going beyond perturbation theory the soft function,
which we identify as the longitudinal momentum distribution in an
on-shell b quark, is replaced by the b--quark distribution in the B meson.
The two differ by power corrections.
We show how the resummation of running--coupling effects can be used
to perform consistent separation to power accuracy between perturbative
and non-perturbative contributions. In particular, we prove that the
leading infrared renormalon ambiguity in the Sudakov exponent cancels
against the one associated with the definition of the pole mass.
This cancellation allows us to identify the non-perturbative parameter
that controls the shift of the perturbative spectrum in the heavy--quark
limit as the mass difference between the meson and the quark. }


\preprint{Cavendish-HEP-04/12}

\begin{document}

\section{Introduction}

Inclusive B--meson decay processes, such as the semi-leptonic
process $\bar B\longrightarrow X_u l^-{\bar{\nu}}_l$ and the rare
$\bar B\longrightarrow X_s\gamma$ process, attract much attention in
the recent years \cite{Grinstein:vj,Jezabek:1988ja,Chay:1990da,Ali:1990tj,
Bigi:1992su,Bigi:1993fe,Blok:1993va,
Manohar:1993qn,Neubert:1993ch,Falk:1993vb,
Neubert:1993um,Bigi:1993ex,Mannel:1994pm,Korchemsky:1994jb,
Kapustin:1995nr,Akhoury:1995fp,Kagan:1998ym,Ligeti:1999ea,DeFazio:1999sv,LR,LLR,
Bauer:2000ew,Neubert:2001sk,Leibovich:2001ra,Bigi:2002qq,
Bauer:2003pi,Bosch:2004th}. This is primarily because of the
availability of increasingly accurate experimental data and the
role of these decays in determining the parameters of the
Cabbibo-Kobayashi-Maskawa (CKM) matrix as well as their potential
in accessing physics beyond the standard model. Owing to
experimental constraints, measurement are usually restricted to
certain kinematical regions so the theoretical prediction of
differential decay distributions is invaluable.

The inclusive nature of the measurement and the fact that the
$b$--quark mass ($m$) is large compared to the QCD
scale~($\Lambda$) guarantee infrared safety. Consequently, the
b--quark decay spectra are calculable in perturbative QCD. On the
other hand, the fact that the b quark is part of a bound state has
a significant impact on the spectra. It has been understood long
ago that the relevant non-perturbative information is contained in
the longitudinal--momentum distribution of the b quark in the B
meson~\cite{Mannel:1994pm}. However, the limited understanding of
the properties of this distribution has been a major stumbling
block in developing precision phenomenology.

The theoretical description of the spectra is particularly
challenging near the endpoints, i.e. near maximal photon energy in
$\bar B\longrightarrow X_s \gamma$, or near maximal charged--lepton
energy in semi-leptonic decays.  Describing the spectra in terms
of $x\equiv 2E/M$, where $E$ is the photon (or charged--lepton)
energy in the B rest frame and $M$ is the meson mass, the endpoint
corresponds to the $x \longrightarrow 1$ limit. It turns out that
the $x\sim 1$ region is experimentally important. This is so, in
particular, for the $\bar B\longrightarrow X_u l^-{\bar{\nu}}_l$ decay
which can be accurately measured only beyond the $\bar B\longrightarrow
X_c l^-{\bar{\nu}}_l$ threshold.

Kinematic constrains imply that near the endpoints the hadronic
final state has large energy, ${\cal O}(M)$, but small invariant
mass, ${\cal O}(M\sqrt{1-x})$. Furthermore, radiation off the
heavy quark has typical transverse momenta of order $M(1-x)$.
Since the quark is confined in a meson this radiation is
physically inseparable from interaction with the light degrees of
freedom in the meson. Thus, the binding of the heavy quark, its
kinetic energy as well as other properties of the bound state are
probed at large $x$ where $1-x\sim{\cal O}(\Lambda/M)$.

At the perturbative level, the constrained real--emission phase
space at large $x$ is reflected in appearance of Sudakov logs,
$\ln (1-x)$. Such large corrections appear owing to the
singularity of the parton--branching probability in the soft and
the collinear limits. As a consequence, any fixed--order
perturbative result diverges in the $x\longrightarrow 1$ limit
although the physical distribution vanishes there. All--order
resummation of the logarithmically--enhanced terms is therefore
essential to recover the qualitative behaviour of the distribution
near the endpoint. Sudakov resummation can indeed be performed
using the standard tools of QCD factorization
\cite{Sterman:1986aj,CSS,CT,Catani:1992ua,Korchemsky:1992xv,CLS,CMW,Akhoury:1998gs},
which is based on identifying within the multi-scale physical
distribution separate subprocesses each described by a function of
a single physical scale. Specifically, in inclusive decays there
are three subprocesses: hard, jet and soft that are described by
functions of $M$, $M\sqrt{1-x}$ and $M(1-x)$, respectively.
Resummation of the logarithms is then achieved by solving
perturbative evolution equations for these functions~\cite{Korchemsky:1994jb}.

It is clear, however, that enhanced corrections appear also
non-perturbatively. While for moderate $x$ power corrections
appear on the scale $M$, which is significantly larger than the
QCD scale $\Lambda$, at large $x$ there are enhanced corrections
depending on the jet mass scale, $M\sqrt{1-x}$, and, most
importantly, on the soft scale, $M(1-x)$. Since, in the region of
interest, the latter is of the order of the QCD scale, one deals
here with an infinite set of power corrections, which can be
summed up into a non-perturbative shape function~\cite{Neubert:1993um,Korchemsky:1994jb}.

Much theoretical work has already been done on inclusive B--decay
spectra~\cite{Grinstein:vj,Jezabek:1988ja,Chay:1990da,Ali:1990tj,
Bigi:1992su,Bigi:1993fe,Blok:1993va,
Manohar:1993qn,Neubert:1993ch,Falk:1993vb,
Neubert:1993um,Bigi:1993ex,Mannel:1994pm,Korchemsky:1994jb,
Kapustin:1995nr,Akhoury:1995fp,Kagan:1998ym,Ligeti:1999ea,DeFazio:1999sv,LR,LLR,
Bauer:2000ew,Neubert:2001sk,Leibovich:2001ra,Bigi:2002qq,
Bauer:2003pi,Bosch:2004th}. Most recent
investigations~\cite{Bauer:2003pi,Bosch:2004th} have been based on
the so-called soft collinear effective
theory~\cite{Bauer:2000yr,Bauer:2001yt,Chay:2002vy,Beneke:2002ph,Hill:2002vw},
by which factorization is implemented at the level of fields.
Despite this progress there are certain theoretical questions that
remain open and limit the prospects of precision phenomenology.
This includes, in particular, the question of how to
systematically separate between perturbative and non-perturbative
corrections on the soft scale. The present work is a step in
resolving this question.

The crucial observation is that for infrared and collinear safe 
observables in general, and for B--decay spectra in particular, 
the separation between perturbative and non-perturbative contributions 
should be based on {\em power accuracy}. This stands in contrast with
the commonly used collinear factorization, which applies for example to 
structure functions in deep inelastic scattering, where the separation is based 
on logarithmic accuracy.
In fact,  when considering the decay spectrum near the endpoint, there is no 
effective way to perform logarithmic separation. Such
separation would have only been effective had there been a large
gap between $M(1-x)$ and $\Lambda$ such that one could have chosen
a factorization scale $\mu$ which satisfies $\Lambda \ll \mu\ll
M(1-x)$. In the region of interest the soft function
receives perturbative and non-perturbative contributions from
roughly the same momentum scales. The perturbative contributions
correspond to gluon radiation off the heavy quark, radiation that
peaks at transverse momentum $k_{\perp}\sim m(1-x)$. The
perturbative soft function describes the longitudinal--momentum
distribution of a (slightly off-shell) b quark inside an on-shell
b quark. The non-perturbative contribution amounts to
power--suppressed corrections which distinguish the latter from
the b--quark distribution inside a B meson. Note that although the
on-shell b quark is not an asymptotic state --- the full quark
propagator has no pole --- the b--quark distribution in an
on-shell b quark is well defined to  any order in perturbation
theory\footnote{This object is infrared and collinear safe, so the
only singularities which do not cancel between real-emission and
virtual diagrams are ultraviolet (singularities that cancel out
when the soft function is combined with the other subprocesses
building up the observable decay spectrum).}. The price one
eventually pays for dealing with a non-physical object, the
on-shell quark, is parametrically small: it is the
power--suppressed ambiguity in the sum of perturbation theory.

This distinction between the QCD analysis of infrared and
collinear safe observables and the standard, collinear
factorization can be formulated also in terms of evolution. 
Collinear factorization is based on computing the anomalous dimension
perturbatively and then supplementing the evolution equation by a
purely non-perturbative initial condition. On the other hand for
infrared and collinear safe observables one computes
perturbatively also the initial condition and the role of
non-perturbative corrections is restricted to modifying this
initial condition by power terms.

Our plan is to extend the concept of large--$x$ factorization
and the resulting Sudakov resummation beyond any logarithmic accuracy.
In order to explain how this is done, let us first recall
the standard implementation of Sudakov resummation, e.g.~\cite{Akhoury:1995fp,LR,LLR},
and examine its limitations. It is worthwhile stressing already now
that these limitations are not the consequence of a specific mathematical
procedure, but instead, they are inherent to the logarithmic accuracy criterion. 

It is convenient to formulate Sudakov resummation
in moment space; see \eq{MN} below for a definition. The moment
index $N$ is the Mellin conjugate variable to~$x$, so large
moments become increasingly sensitive to the interesting
$x\longrightarrow 1$ limit. Owing to momentum conservation
contributions of different subprocesses at large $x$ are simply
written as a product in moment space. Furthermore, the
factorization property of soft and collinear matrix elements  
implies that the soft subprocess (and likewise the jet) can be
written in moment space as an exponential. 
The standard approach
to Sudakov resummation is based on computing the moment--space
exponent to a fixed logarithmic accuracy, usually
next--to--leading logarithmic accuracy, thus resumming terms of
the form $\alpha_s^n\ln^k N$ where $k=n+1$ (leading logs --- LL)
as well as $k=n$ (next--to--leading logs --- NLL) in the exponent.
This contribution can be summed over $n$ to all orders, yielding a
well--defined function. Nevertheless, this function has unphysical
poles (Landau singularities) at $N\sim {\cal O} (m/\Lambda)$. When
converting the resummed series back to $x$ space these poles lead
to large fluctuations at large~$x$. The perturbative result is
then unstable with respect to higher--order
corrections\footnote{Moreover, the Landau pole makes the Mellin
integral ambiguous (integration--contour dependent). This
ambiguity should not be confused with the renormalon ambiguity (to
be discussed below) which presents itself already when resumming
the series in moment space. Nevertheless, it does provide an
indication of the inapplicability of perturbation theory in the
region $m(1-x)\sim \Lambda$.} for $x$ values between the peak and
the endpoint.

While the moment--space perturbative result is well defined at any given 
logarithmic accuracy, the corrections owing to subleading
logarithms (that are neglected) are large. Because of the
sensitivity of the contributing diagrams to small momenta,
subleading logarithms $k<n$ have larger numerical coefficients
which eventually break the hierarchy on which the perturbative expansion is
based. This is a result of infrared renormalon contributions to
the Sudakov exponent~\cite{Gardi:2001ny}. Similarly to other
perturbative series in QCD~\cite{Dokshitzer:1995qm,ren_review},
owing to infrared renormalons the Sudakov exponent is
non-Borel-summable. Upon summing the perturbative expansion 
{\em with no restriction on the logarithmic accuracy} (i.e. on $n-k$) the exponent
develops power--suppressed ambiguities. In this way resummation of
running--coupling effects leads to identifying
power--like ambiguities in the moment--space exponent opening up
the possibility to perform separation at power accuracy between
perturbative and non-perturbative corrections.

The appearance of renormalons in the Sudakov exponent is not
special to B--decay spectra but is rather common to a wide range
of QCD distributions near kinematic thresholds. The Dressed Gluon
Exponentiation (DGE) approach deals with this situation by
performing a renormalon calculation of the Sudakov exponent,
defining it by an appropriate regularization prescription, such as
the principal value of the Borel integral, and finally
parametrizing non-perturbative corrections to the moment--space
exponent based on the parametric form of the ambiguity. It is
worth noting that, contrary to any fixed--logarithmic--accuracy
resummation, the principal--value regularization of the Sudakov
exponent has no Landau singularities. Existing applications of DGE range
from event--shape distributions in the two--jet
limit~\cite{Gardi:2001ny,Gardi:2003iv}, through deep inelastic
structure functions at large
Bjorken~$x$~\cite{Gardi:2001di,Gardi:2002bk,Gardi:2002xm}, to the
heavy--quark fragmentation function~\cite{CG} which is closely
related to B decay.

Aiming at power accuracy we begin our perturbative 
analysis of decay spectra by performing resumming of running--coupling effects. 
In practice the calculation is done, as in other applications, with a single
dressed gluon, i.e. to leading order in the large--$\beta_0$ limit. 
We exponentiate the $x\longrightarrow 1$ singular terms in moment space, 
identify the renormalon ambiguity of the Sudakov exponent and {\em define} the sum of the series 
by the principal value of the Borel integral, a choice that also defines the
non-perturbative correction. 

In the b--quark distribution analyzed here, much like the
heavy--quark fragmentation function~\cite{Nason:1996pk,CG,JR}, the
ambiguity of the Sudakov exponent corresponds to powers of
$N\Lambda/m$. The ambiguity implies the
existence of genuine non-perturbative corrections to the exponent
of a similar parametric form. Independently of how these power
corrections are parametrized, as soon as they become important the
renormalon ambiguities of the Sudakov exponent cannot be ignored.
From this perspective it is hard to interpret any phenomenological
description of the non-perturbative b--quark distribution in the 
B meson which does not take into account the renormalons.

Deeper understanding the b-quark distribution function 
requires establishing the cancellation of ambiguities between 
its perturbative and non-perturbative ingredients. 
This separation is ambiguous because it is based on the concept 
of an on-shell heavy quark.
Using the heavy quark effective theory (HQET) \cite{Neubert:1993mb,Neubert:1996wg} 
we show that the leading power correction, of order $N\Lambda/M$, 
cancels between the perturbative Sudakov exponent and non-perturbative 
corrections involving the quark pole 
mass~\cite{Falk:1992fm,Bigi:1994em,Beneke:1994sw,BBZ,Neubert:1994wq,Bigi:1996si}.
Defining both the Sudakov exponent and the pole mass by the
same prescription (e.g. principal value of the Borel sum) the parameter that controls
the corresponding non-perturbative effect on the spectrum --- a
shift --- is uniquely identified. The final result is independent
of the prescription used. This extends the previous result on cancellation 
of renormalon ambiguities in the total decay width~\cite{BBZ} to the 
level of differential decay spectra.

The structure of this paper is as follows. In section~\ref{BXsG}
we consider the radiative $\bar B\longrightarrow X_s\gamma$ decay
spectrum, present the perturbative DGE result for the moments and
then consider in some detail the power--like separation between
perturbative and non-perturbative contributions to the b--quark
distribution. In particular, we identify the relevant non-perturbative 
parameters using the HQET and prove the cancellation of the first infrared
renormalon ambiguity. Although we refer specifically to the
radiative decay, the analysis of the b--quark distribution is
general: it applies to other inclusive decays as well. In section
\ref{SL_decay}, we turn to the semi-leptonic decay spectrum
focusing on the analogy and differences with the radiative decay.
Appendix \ref{App_sl} gives some more details of the corresponding
renormalon calculation where the soft and jet functions are
computed separately. Section~\ref{conc} is reserved for
conclusions.

\section{$\bar B\longrightarrow X_s\gamma$ decay\label{BXsG}}

In this section we address perturbative and non-perturbative
aspects of the energy spectrum in the radiative B decay,
$\bar B\longrightarrow X_s\gamma$. We begin by computing the photon
energy distribution to all orders in the large--$\beta_0$ limit.
This is done by performing the one--loop calculation with a
Borel--modified gluon propagator. Next we focus on the endpoint
region, first in perturbation theory and then beyond. In
section~\ref{A_renormalon_calc} we compute the Sudakov exponent in
moment space. We then show (section~\ref{SJH}) that this result is
consistent with factorization into hard, jet and soft functions,
and match the DGE result with
NLL accuracy and with the full ${\cal
O}(\alpha_s)$ result (section~\ref{Matching_sec}). To end the
perturbative part we demonstrate in section~\ref{FLA_PV} the
divergence of the Sudakov exponent and analyse the difference
between truncation at fixed logarithmic accuracy and
principal--value resummation. Non-perturbative aspects are
discussed in section \ref{PC}, beginning, in
section~\ref{The_shape_function}, with precise identification of
the relevant power corrections (the shape function) through the moment--space ratio
between the quark distribution in a meson and the one in an
on-shell heavy quark. In section~\ref{HQET} we identify the first few 
parameters that control power corrections in the heavy--quark limit 
using the HQET and finally, in section~\ref{cancellation}, we 
prove the cancellation of the leading renormalon ambiguity.

Let us first define the kinematics. We denote the inclusive
differential $\bar B\longrightarrow X_s\gamma$ decay rate by $\Gamma$,
and define
\[x\equiv 2p_{\gamma}\cdot p_B/M^2,\]
where $p_{\gamma}$ and $p_B$ are the photon and B--meson momentum,
respectively ($p_{\gamma}^2=0$, $p_{B}^2=M^2$). In the $B$ rest
frame $x$ represents the photon energy fraction:
$x=2E_{\gamma}/M$.

The short--distance electroweak physics responsible for the decay 
can be integrated out yielding an effective Hamiltonian description 
which involves eight operators of dimension five and six~\cite{Grinstein:vj}. 
Calculation of the corresponding coefficient functions, the 
evolution from $W$--mass scale to the $b$--quark mass scale,
and perturbative evaluation of the matrix elements
shows that the $\bar B\longrightarrow X_s\gamma$ decay is 
dominated by the following magnetic operator~\cite{Ali:1990tj,Kapustin:1995nr,Ligeti:1999ea}:
\begin{equation}
H=\sqrt{2} G_F V_{tb}V_{ts}^*A(m_t^2/m_W^2)\, \frac{e m}{16\pi^2}\,
 \bar{s}_L\sigma^{\mu \nu}b_R F_{\mu \nu},
\label{O7}
\end{equation}
where $F_{\mu \nu}$ is the photon field strength. The contribution of 
this particular operator is singled out by being singular in the 
$x\longrightarrow 1$ limit in which we are interested. 
Therefore, in the following we only consider this operator.
We would like to emphasise that a complete analysis needed 
for phenomenology would have to
take into account other operators\footnote{It has been shown~\cite{Ali:1990tj,Kapustin:1995nr,Ligeti:1999ea} that 
the contribution of operators other than \eq{O7} is suppressed 
at large $x$ by a power of $(1-x)$, making it unimportant for
$x\gsim 0.9$. In the region
$0.9 \gsim x\gsim 0.7$ these operators lead to moderate 
corrections; see e.g. Fig.~2 in \cite{Ligeti:1999ea}. At 
smaller--$x$ values (the tail of the spectrum) the magnetic 
operator of \eq{O7} ceases to dominate.} as well. There, 
however, the standard perturbative treatment is likely 
to be sufficient.

\subsection{Perturbation theory}

\subsubsection{A renormalon calculation\label{A_renormalon_calc}}

Referring to the $B$ meson as if it were a free on-shell b quark, 
the inclusive distribution in $x$  is infrared and 
collinear safe, so it can be computed perturbatively in QCD. 
At the perturbative level the
scaling variable $x$ becomes $2p_{\gamma}\cdot p/m^2$ where $p$ is
the b--quark momentum and $m$ is the pole mass, $p^2=m^2$.
A calculation of the real--emission cuts
with a single dressed gluon yields the following
distribution:
\begin{eqnarray}
\label{BXg_full}
\frac{1}{\Gamma_{\tot}}\frac{d\Gamma}{dx}&=&(1+{\cal O}(\alpha_s))\delta(1-x)
+\frac{C_F}{2\beta_0}
\int_0^{\infty}du\,T(u)\,\left(\frac{\Lambda^2}{m^2}\right)^u\,B(x,u)
\end{eqnarray}
with
\begin{eqnarray}
\label{B_BXSG}
 B(x,u)&\equiv&{\rm e}^{cu}\,\frac{\sin\pi u}{\pi u}
\,x^3\,(1-x)^{-u} \int_0^1 d\alpha\,
\alpha\,(1-\alpha)^{-u}\,\times \\ \nonumber &&\hspace*{-30pt}
\bigg[\frac{1}{(1-x\alpha)^2}\left(1-4\alpha+\alpha^2
-\frac{(1-\alpha)^2}{1-u}\right)
+\frac{1-\alpha}{(1-x\alpha)(1-u)} +\frac
2{1-x}\frac{1}{1-x\alpha} +\frac 1{1-x}\bigg],
\end{eqnarray}
where $\Lambda$ is in the ${\overline{\rm MS}}$ scheme and
$c=5/3$.

\begin{figure}[th]
\begin{center}
\epsfig{file=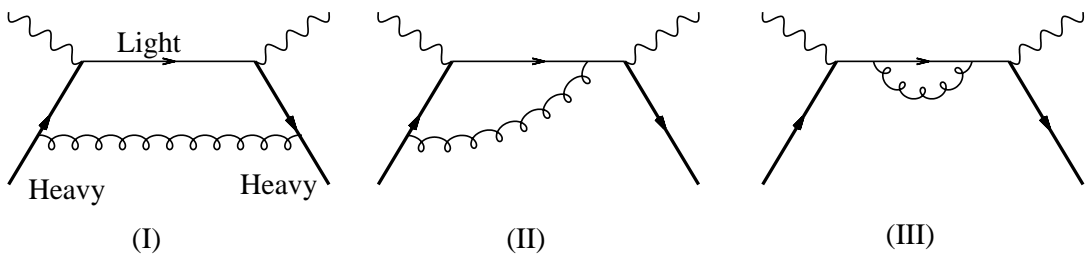,angle=0,width=14.4cm}
\caption{\label{fig:diagrams} The three one--loop diagrams contributing to
the squared decay matrix element of a heavy quark (thick
line) into a light quark (horizontal, thin line).}
\end{center}
\end{figure}

To arrive at this result we computed the imaginary part of the
three one--loop diagrams of Fig.~\ref{fig:diagrams} with a
Borel--modified gluon propagator,
\[
\frac{1}{-k^2}\longrightarrow \frac{(\Lambda^2)^u}{(-k^2)^{1+u}}.
\]
Integrating over the Borel parameter $u$ is equivalent to
resumming a gauge--invariant set of diagrams with a single dressed
gluon to all orders. The set of diagrams is defined by evaluating
the running coupling at the scale of the gluon virtuality,
\begin{equation}
  \frac{\beta_0\alpha_s(-k^2)}{\pi}=\int_0^\infty du \,T(u)
  \,\left(\frac{\Lambda^2}{-k^2}\right)^{u},
  \label{Borel_def}
\end{equation}
and it reduces to the insertion of an arbitrary number of fermion
loops in the large--$N_f$ limit; see e.g. \cite{ren_review}. Here
$\beta_0$ is the leading--order coefficient of the $\beta$
function,
\begin{equation}
\frac{d(\alpha_s/\pi)}{d\ln\mu^2}=-\beta_0(\alpha_s/\pi)^2-\beta_1(\alpha_s/\pi)^3+\ldots
\hspace*{60pt}\beta_0=\frac{11}{12}C_A-\frac16 N_f\,,
\label{beta_0}
\end{equation}
and the exponential factor ${\rm e}^{c u}$ in \eq{B_BXSG}
originates in the renormalization of the fermion loop in the
$\overline {\rm{MS}}$ scheme. Throughout this paper we shall be
using the scheme--invariant formulation~\cite{Grunberg} of the
Borel transform. In the large--$\beta_0$ limit (one--loop running
coupling) $T(u)\equiv 1$. This function is introduced here so that
the running coupling can be considered beyond one loop, e.g.
truncating \eq{beta_0} at two loops one has:
\begin{equation}
T(u)=\frac{(u\delta)^{u\delta}{\rm
e}^{-u\delta}}{\Gamma(1+u\delta)}. \label{Tu}
\end{equation}
where $\delta=\beta_1/\beta_0^2$

In \eq{B_BXSG} the terms in the square brackets are written as
they appear from individual diagrams in the Feynman gauge. The
first two terms corresponds to gluon emission from the b quark
prior to its decay (diagram I), the last (fourth) term to gluon
emission from the strange quark in the final state (diagram III)
and the third term to interference between the two amplitudes
(diagram II).

Upon expanding the Borel integrand near $u=0$ to leading order
(${\cal O}(u^0)$), \eq{BXg_full} reproduces the
known~${\cal O}(\alpha_s)$ real--emission contribution to this
process ({\em cf.}~\cite{Kapustin:1995nr}),
\begin{eqnarray}
\label{BXg_LO_x}
&&\frac{1}{\Gamma_{\tot}}\frac{d\Gamma}{dx}=(1+{\cal O}(\alpha_s))\delta(1-x)\nonumber \\
&&\quad +\frac{C_F\alpha_s}{2\pi}\left[
-\left(1+x+\frac{2}{1-x}\right)\ln(1-x)-\frac72\frac{1}{1-x}+\frac72
+\frac12x-x^2\right].
\end{eqnarray}
Beyond this order \eq{BXg_full} resums running--coupling effects
to all orders in perturbation theory in a scheme invariant manner.
In particular, renormalization--scale dependence is avoided. With
$T(u)=1$ this resummation comprises at any order $n$ of the terms
of the form $C_F\beta_0^n\alpha_s^{n+1}$. Using $T(u)$ this
resummation is generalized to include the effects of subleading
corrections to the running of the coupling.

\subsubsection{Borel representation of the Sudakov exponent\label{Borel_rep_}}

The most striking feature of the $\bar B\longrightarrow X_s\gamma$
distribution is that it peaks close to the end--point $x\sim 1$,
and yet vanishes at $x=1$. In perturbation theory this is a
consequence of Sudakov suppression~\cite{Korchemsky:1994jb}. At
Born level the distribution is just a delta function at $x=1$, so
the main effect of subleading corrections is to smear this peak.

As \eq{BXg_LO_x} shows, singular terms at~$x\longrightarrow 1$
appear in perturbation theory from both real emission and virtual
corrections. The latter are proportional to $\delta(1-x)$. Real
and virtual terms are related in such a way that upon averaging
the distribution (with any finite resolution) a finite result is
obtained. In practice it is most useful to consider the normalized
moments
\begin{equation}
M_N\equiv \int_0^1
dx\,\frac{1}{\Gamma_{\tot}}\frac{d\Gamma}{dx}x^{N-1}, \quad \quad
M_1\equiv 1, \label{MN}
\end{equation}
because in moment space factorization of contributions from
different parts of phase space takes the form of a product.

In order to compute the Sudakov exponent we first extract from
\eq{BXg_full} the divergent (logarithmically enhanced) terms ---
see Appendix \ref{App_BXSG} --- obtaining:
\begin{eqnarray}
\label{BXg_logs_fact} \frac{1}{\Gamma_{\tot}}\frac{d\Gamma}{dx}
&=&(1+{\cal O}(\alpha_s))\delta(1-x) +\frac{C_F}{2\beta_0}\int_0^{\infty}du
\,T(u)\,\left(\frac{\Lambda^2}{m^2}\right)^u\,{\rm e}^{cu}\,\times
\\\nonumber&&\qquad \bigg[ \frac 2u (1-u)\,(1-x)^{-1-2u}  -
\left(\frac2u+\frac{1}{1-u}+\frac{1}{2-u}\right) \frac{\sin\pi
u}{\pi u}\,(1-x)^{-1-u}\bigg].
\end{eqnarray}
The corresponding perturbative sum contains all the
$x\longrightarrow 1$ singular terms which are associated with the
running coupling in the single--dressed--gluon approximation, to
{\em any logarithmic accuracy}. In \eq{BXg_logs_fact} we clearly
identify the two scales which were anticipated in the
introduction: the soft scale $m^2(1-x)^2$, corresponding to
typical transverse momentum of gluon radiation off the heavy
quark, and the jet scale $m^2(1-x)$, corresponding to the
invariant mass of the unresolved jet in the final state. Dynamics
on both these scales generate Sudakov logs.

Next we consider the photon energy moments defined by \eq{MN}.
Sudakov logarithms exponentiate in {\em moment space} as:
\begin{eqnarray}
\label{BXg_logs_N_2loop} &&M_N^{\PT}=C_N(\alpha_s(m^2))\,\exp
\bigg\{ \frac{C_F}{2\beta_0}\int_0^{\infty}du \,T(u)\,
\left(\frac{\Lambda^2}{m^2}\right)^u\,{\rm e}^{cu}\,\times\\
\nonumber &&\hspace*{30pt} \bigg[ \frac 2u
(1-u)\,\Gamma(-2u)\left(N^{2u}-1\right)
 -
\left(\frac2u+\frac{1}{1-u}+\frac{1}{2-u}\right) \frac{\sin\pi
u}{\pi u}\,\Gamma(-u)\left(N^{u}-1\right)\bigg] \bigg\},
\end{eqnarray}
where $C_N(\alpha_s(m^2))$ represents contributions which are
finite for $N\longrightarrow \infty$. Contributions to the
Sudakov exponent associated with the soft and the jet
subprocesses are easily distinguished based on their
parametric dependence on $N$.

The exponentiation of logarithms to any logarithmic accuracy
unavoidably implies that power corrections also exponentiate:
infrared renormalons in the Sudakov exponent appear, as usual, upon
integrating over $x$ near the endpoint. This occurs in both
subprocesses: power corrections on the scale $m(1-x)\sim m/N$ are
related to the radiation that accompanies a heavy quark, while
those on $m^2(1-x)\sim m^2/N$ to the formation of the jet in the
final state. Thus, power suppressed ambiguities scaling as integer
powers of $N\Lambda/m$ and $N\Lambda^2/m^2$ are part of the
perturbative Sudakov exponent. These power--suppressed ambiguities
must cancel with corresponding ambiguities in the non-perturbative
function. This suggests that the physical moments of~\eq{MN}
can be expressed as
\begin{equation}
M_N=M_N^{\PT} \times M_N^{\NP},
\label{PT_NP_sep}
\end{equation}
and that  the function $M_N^{\NP}$ itself takes a form
which is similar to the ambiguities\footnote{This idea is the basis
for the renormalon approach to power corrections~\cite{Dokshitzer:1995qm,ren_review}.}.
Reading in \eq{BXg_logs_N_2loop} the residues of the soft scale,
one can write an ansatz for the non-perturbative soft function
without even considering its physical origin:
\begin{eqnarray}
\label{BXg_powers} &&M_N^{\NP}=\exp
\bigg\{-\epsilon_1\frac{(N-1)\Lambda}{M}-\epsilon_3
\left(\frac{(N-1)\Lambda}{M}\right)^3-\epsilon_4 \left(\frac{(N-1)\Lambda}{M}\right)^4 \cdots
 \bigg\},
\end{eqnarray}
where we introduced a non-perturbative (at this point ambiguous)
parameter $\epsilon_i={\cal O}(1)$ for each Borel pole in the
perturbative exponent. Note the absence of the second power of
$(N-1)\Lambda/{M}$ in this ansatz owing to the vanishing of the
corresponding residue. As in the perturbative Sudakov exponent we
neglected here terms that are suppressed by a relative power of
$1/N$ (or, equivalently, $\Lambda/M$). In section \ref{PC} we
shall return to consider power corrections on the soft scale from
a different perspective.

\subsubsection{Factorization into soft, jet and hard subprocesses\label{SJH}}

Exponentiation is a direct consequence of factorization, and it
can be derived (see e.g.~\cite{CLS}) as the solution of
renormalization group equations for the separate subprocesses.
Indeed \eq{BXg_logs_N_2loop} is consistent with the general
factorization formula of Ref.~\cite{Korchemsky:1994jb}. The
dependence of the different terms in \eq{BXg_logs_N_2loop} on $N$
suggest the following separation:
\begin{eqnarray}
M_N^{\PT}=C_N(\alpha_s(m^2))\,S_N(m^2;\mu^2)\,J_N(m^2;\mu^2).
\label{gen_fact_form}
\end{eqnarray}
where
\begin{eqnarray}
\label{Soft}
S_N(m^2;\mu^2)&=&\exp \bigg\{
\frac{C_F}{\beta_0}\int_0^{\infty}\frac{du}{u} \,T(u)\,
\left(\frac{\Lambda^2}{m^2}\right)^u\,\times\\
\nonumber &&\hspace*{30pt} \bigg[
B_S(u) \Gamma(-2u)\left(N^{2u}-1\right)
 + \left(\frac{m^2}{\mu^2}\right)^u
B_{\cal A}(u)\ln N \bigg] \bigg\},
\end{eqnarray}
with
\begin{equation}
B_S(u)={\rm e}^{cu} \,(1-u)\,+{\cal O}(1/\beta_0),
\end{equation}
and
\begin{eqnarray}
\label{Jet}
J_N(m^2;\mu^2)&=&\exp \bigg\{
\frac{C_F}{\beta_0}\int_0^{\infty}\frac{du}{u} \,T(u)\,
\left(\frac{\Lambda^2}{m^2}\right)^u\,\times\\
\nonumber &&\hspace*{-30pt} \bigg[ -\,B_J(u) \Gamma(-u)\left(N^{u}-1\right)
-\left(\frac{m^2}{\mu^2}\right)^u
B_{\cal A}(u)\ln N
 \bigg] \bigg\},
\end{eqnarray}
with
\begin{equation}
B_J(u)=\frac12 {\rm e}^{cu}
\left(\frac{1}{1-u}+\frac{2}{2-u}\right) \frac{\sin\pi
u}{\pi u} +{\cal O}(1/\beta_0).
\end{equation}
Here $C_N(\alpha_s(m^2))$ is the hard function, $S_N(m^2;\mu^2)$
 the soft function and
$J_N(m^2;\mu^2)$ the jet function.
As usual, the price of the separation into subprocesses is the
dependence of each of the functions on the factorization scheme
and scale ($\mu^2$).
To make the soft and the jet functions well defined
(the corresponding integrand free of $1/u$ singularity)
we have added and subtracted a term proportional to
\[B_{\cal A}(u)\ln N=(1+cu+{\cal O}(u^2))\ln N,\]
which is the Borel representation of the cusp anomalous dimension
\cite{Korchemsky:1988si,Korchemsky:1992xv}.
This function is free of renormalons in MS-like schemes.
The resulting dependence on the scale is
\begin{eqnarray}
\label{evolution_SJ}
-\frac{d \ln S_N(m^2;\mu^2)}{d\ln \mu^2}=
\frac{d \ln J_N(m^2;\mu^2)}{d\ln \mu^2}&=&
\frac{C_F}{\beta_0}\ln N\,\times\,
\int_0^{\infty}du \,T(u)\left(\frac{\mu^2}{\Lambda^2}\right)^u B_{\cal A}(u)\nonumber\\
&\equiv&
\frac{C_F}{\beta_0}\ln N \,\times\,\Gamma_{\cusp}(\alpha_s(\mu^2)) .
\end{eqnarray}

The specific factorization we made is natural because it implies
normalization of the separate soft and jet functions to unity,
$S_{N=1}(m^2;\mu^2)=1$ and $J_{N=1}(m^2;\mu^2)=1$, independently
of the factorization scale $\mu^2$. Related features are reflected
in \eq{evolution_SJ}: (1) the logarithmic derivatives of $\ln S_N$
and $\ln J_N$ with respect to $\mu^2$ depend on $\mu^2$ and on the
QCD scale $\Lambda^2$ only through the coupling; and (2) the hard
function does not depend on the factorization scale.

Factorization similar to \eq{gen_fact_form} was already introduced
in~\cite{Gardi:2002xm,CG} in the application of DGE to other
physical processes. In fact, Eq.~(\ref{BXg_logs_N_2loop}) as a
whole is {\em identical} to that of the single heavy hadron
inclusive cross section in $e^+e^-$ annihilation, eq. (60) in
\cite{CG}, upon replacing the center--of--mass energy $q$, which sets the scale 
of hard and jet functions in the production process, by $m$.
The relation between the two processes is that of crossing: the b
quark is in the initial state in the decay process while it is in
the final state in the production process of \cite{CG}. The jet
function, \eq{Jet}, is also identical to the one controlling the
large--$x$ limit of deep inelastic structure functions, see Eq.
(15) in~\cite{Gardi:2002xm}.

The unusual feature of the factorization formula
of~\eq{gen_fact_form} is that the soft function as well as the jet
function depend not only on their natural scale, $m/N$ and
$m/\sqrt{N}$, respectively, but also on the hard scale $m$ which
plays the role of a physical ultraviolet cutoff. Strictly speaking
this violates the idea of factorization, but this violation is
unimportant as it does not affect the universality of these
functions. As a consequence, however, the evolution equations of
\eq{evolution_SJ} differ from the ones derived in
Ref.~\cite{Korchemsky:1994jb} based on infrared factorization. For
the same reason the functions $S_N(m^2;\mu^2)$ and
$J_N(m^2;\mu^2)$ do not directly correspond to the functions
defined in the effective theory
approach~\cite{Bauer:2000yr,Bauer:2001yt,
Chay:2002vy,Beneke:2002ph,Hill:2002vw,Bauer:2003pi,Bosch:2004th}.
Strict factorization can be implemented at the price of having
either a $\Lambda$--dependent or a $\mu$--dependent anomalous
dimensions and loosing the natural, $\mu$--independent
normalization of the functions.

\subsubsection{Matching to fixed logarithmic accuracy
and fixed--order calculations\label{Matching_sec}}

The calculation presented in the previous section considered only
running--coupling effects. However, beyond the leading logarithmic
accuracy the Sudakov exponent gets contributions from other
diagrams having different color factors.

In order to comply with full next--to--leading logarithmic
accuracy (NLL) two modifications of the large--$\beta_0$--limit
result are needed. First, as mentioned above the coupling needs to
be promoted to run according to the two--loop $\beta$ function,
\eq{beta_0}. We do this using $T(u)$ of \eq{Tu}.

In addition new NLL terms appear, which are related to the
next--to--leading order term in the cusp anomalous dimension,
affecting both subprocesses in \eq{BXg_logs_N_2loop}:
\begin{equation}
\frac{1}{u} \longrightarrow
\frac{1}{u}\left[1+u\left(\frac13-\frac{\pi^2}{12}\right)\frac{C_A}{\beta_0}
+{\cal O} (u^2)\right].
\end{equation}
These corrections can be included by modifying $c$, which equals
$5/3$ in the large--$\beta_0$ limit, and replacing it by
\begin{equation}
c=\frac53+\left(\frac13-\frac{\pi^2}{12}\right)\frac{C_A}{\beta_0}.
\label{c}
\end{equation}
It should be emphasized that promoting the result to comply with
NLL accuracy in this particular way is not a result of
higher--order renormalon calculation; this manipulation is
arbitrary as far as the functional dependence on $u$ is concerned.
The related theoretical uncertainty concerns both subleading
logarithms (NNLL and beyond) and power corrections. For discussion
of this issue in the context of deep inelastic structure functions
at large Bjorken $x$, see \cite{Gardi:2002xm}. Here we do not
investigate this issue further.

Next, the full, fixed--order ${\cal O}(\alpha_s)$ result can be
matched by determining $C_N(\alpha_s(m^2))$ in
\eq{BXg_logs_N_2loop} to this order. In principle, such matching
can be done at any order. 
For example, following the so-called $\log$--$R$ matching
scheme \cite{Catani:1992ua}, one can write:
\begin{equation}
\ln C_N(\alpha_s(m^2))= \left.\ln M_N^{\PT}\right\vert_{{\cal
O}(\alpha_s)}- \left.\ln M_N^{\PT}\right\vert_{{\cal
O}(\alpha_s),\log},
\end{equation}
where $\left. M_N^{\PT}\right\vert_{{\cal O}(\alpha_s)}$ is the
full ${\cal O}(\alpha_s)$ correction\footnote{This term is
extracted from \eq{BXg_full} by expanding the Borel integrand to
leading order in $u$, obtaining \eq{BXg_LO_x}, and then converting
to moment space using the ``plus'' prescription by which virtual
terms proportional to $\delta(1-x)$ are determined by the
normalization condition $M_1=1$.} which includes
logarithmically--enhanced terms, constant terms as well as terms
that are suppressed at large $N$, while $\left.
M_N^{\PT}\right\vert_{{\cal O}(\alpha_s),\log}$ contains just the
logarithmically--enhanced terms at ${\cal O}(\alpha_s)$ which are
subtracted to avoid double counting with the resummed Sudakov
exponent.
It should be noted that there is
no unique way to perform matching: non-logarithmic terms beyond
the order for which the full fixed--order result is known can
vary. 

The explicit result for $C_N (\alpha_s(m^2))$~is:
\begin{eqnarray}
\label{matching} &&C_N (\alpha_s(m^2))=1+
\frac{C_F\alpha_s}{2\pi}\,\bigg\{\bigg[
\left({\displaystyle \frac {7}{2}}  - {\displaystyle \frac
{1}{N\,(N + 1)}}  + {\displaystyle \frac {2}{N}}
\right)\,\left(\Psi (N) + \gamma_E \right)+\Psi_1(N) -
{\displaystyle \frac {\pi ^{2}}{6}}
 \nonumber \\ &&
\hspace*{70pt}- \left(\Psi (N) + \gamma_E \right)^{2}  - {\displaystyle \frac
{31}{6}}  + {\displaystyle \frac {9}{2\,N}}
 + {\displaystyle \frac {1}{(N + 1)^{2}}}  -
{\displaystyle \frac {1}{N + 2}}  - {\displaystyle \frac {1}{2\,N
 + 2}}  + {\displaystyle \frac {1}{N^{2}}} \bigg] \nonumber
 \\&&\hspace*{70pt}-\,
\bigg[-\ln^2(N)+\left(\frac72-2\gamma_E\right)\ln(N)\bigg] \bigg\}
+{\cal O}\left(\alpha_s^2\right).
\end{eqnarray}
Together with \eq{BXg_logs_N_2loop} (or \eq{gen_fact_form}) this completes the
perturbative calculation of the radiative decay moments.

\subsubsection{Principal value vs. fixed logarithmic accuracy\label{FLA_PV}}

The soft scale, $m/N$, becomes of order of the QCD scale $\Lambda$
for moments $N\sim 20$. This regime is certainly non-perturbative.
However, even at much lower moments the ratio between these scales
is not so large and power corrections can be comparable to
perturbative ones. In these circumstances parametrization
of power corrections on the scale $m/N$ is unavoidable.

It is a general phenomenon in QCD that owing to infrared
renormalons the perturbative sum is ambiguous and consistent
separation of power terms requires definition of this sum. This is
true in particular for the Sudakov exponent, where the renormalon
factorial increase is carried by subleading logarithms
\cite{Gardi:2001ny}.

Sudakov logs are usually computed to fixed logarithmic accuracy.
However, when power corrections on the same scale are important
the logarithmic accuracy criterion becomes inappropriate.
Separating between perturbative and non-perturbative terms
requires then power accuracy. The large--$\beta_0$ calculation
presented above probes the large--order asymptotic behaviour of
the perturbative expansion allowing one to introduce power--like
separation between perturbative and non-perturbative contributions
to the Sudakov exponent.

Let us now return to the moments defined in~\eq{MN}. By
perturbative considerations we deduced that $M_N=M_N^{\PT} \times
M_N^{\NP}$, where the perturbative contribution,
\eq{BXg_logs_N_2loop}, was calculated while the non-perturbative
one, \eq{BXg_powers}, parametrized. Each of the two contributions
suffers from renormalon ambiguities which cancel in the product.
The separation between $M_N^{\PT}$ and $M_N^{\NP}$ is a matter of
convention. We implement this separation by a principal--value
regularization of the Borel integral in \eq{BXg_logs_N_2loop}. The
Borel regularization is equivalent in principle to a hard cutoff
on some Euclidean momentum integral~\cite{Gardi:1999dq} but it is
more convenient in practice.

\begin{figure}[th]
\begin{center}
\epsfig{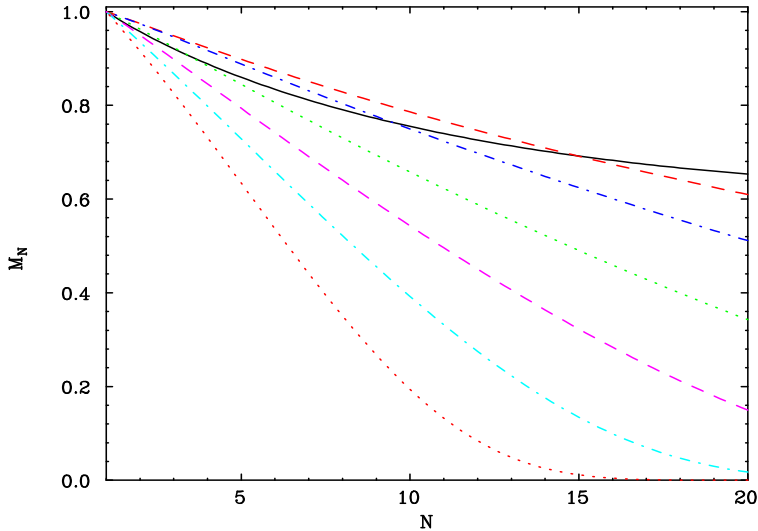}
\caption{\label{fig:nlldge} The (normalized) moments $M_N$ of the
photon spectrum as a function of $N$, as computed through
\eq{BXg_logs_N_2loop}, where $C_N (\alpha_s(m^2))$ is given by
\eq{matching}. The full line is the principal--value Borel sum.
Dashed, dotdash and dotted lines correspond to truncation of the
Sudakov exponent at increasing logarithmic accuracy -- the
uppermost line is LL, then NLL, etc.
}
\end{center}
\end{figure}

In Fig.~\ref{fig:nlldge} the principal-value regularization is
compared with truncation at fixed logarithmic accuracy.
While at low moments the principal--value regularization is
numerically close to NNLL truncation, at higher $N$ it approaches
NLL and then LL, and finally crosses even the LL result developing
milder slope at high $N$. This behaviour is in one--to--one
correspondence with the fact that as $N$ increases, the
perturbative expansion for the exponent breaks down earlier. It
should be emphasized that when the exponent is written at fixed
logarithmic accuracy (NLL and beyond) it suffers from Landau
singularities. This is not the case for the principal--value Borel
sum \cite{CG} which extrapolates smoothly to high--$N$ values.

Note that the NNLL and further subleading logarithms are not
computed exactly but are only estimated based on the expansion of
\eq{BXg_logs_N_2loop}. Nevertheless, we expect that the
corresponding curves in Fig.~\ref{fig:nlldge} represent well the
divergence of the perturbative expansion. As usual in asymptotic
expansions there is no use in summing up the series beyond the
point where it starts diverging, which is roughly the order that
reproduces the principal--value result. Since this never occurs
beyond NNLL an exact calculation of subleading logarithms (yet
unavailable) is less important for phenomenology in this case than
careful treatment of power corrections. What is expected to make a
significant impact on phenomenology is the use of power--like
separation as provided by the principal--value prescription
instead of the commonly used NLL truncation of the Sudakov
exponent. As Fig.~\ref{fig:nlldge} shows, NLL truncation cannot be
considered a good approximation to the principal--value
prescription over the entire $N$ range nor does it differ from it
by power terms.

\subsection{Power corrections\label{PC}}

\subsubsection{The shape function\label{The_shape_function}}

As announced, we focus here on the end--point region, namely the
large--$N$ limit. The leading non-perturbative corrections in this
limit are associated with the soft scale, $m/N$. Through the
renormalon calculation we have provided evidence for power corrections
on the soft scale, but we have not
proven that these are indeed the leading corrections at large~$N$. In fact, they are. 
For example, there are no corrections that scale like $N^2 \Lambda /m$.
A completely general argument will be provided below.

As reflected in the renormalon ambiguity of $S_N(m^2,\mu^2)$ in
\eq{Soft},  the leading power correction is ${\cal O}(N\Lambda/m)$,
which is potentially very large. It is therefore of primary
interest to understand the origin of this correction. This is the
main goal of this section. Subleading corrections on the soft
scale are important as well and need to be
resummed into a shape
function~\cite{Neubert:1993um,Korchemsky:1994jb}, as we discuss
below. On the other hand, we shall not be interested here in
further subleading power corrections ${\cal O}(N\Lambda^2/m^2)$ or
${\cal O}(\Lambda/m)$ which are associated with the jet function
and the hard function, respectively.

To include power corrections on the soft scale we replace
$S_N(m^2;\mu^2)$ in our expression for the moments,
\eq{gen_fact_form}, by the corresponding matrix--element
definition, $\tilde{F}(N;\mu^2)$. Let $B(p_B)$ be a meson with
momentum $p_B$. We define $F(z;\mu^2)$ as the Fourier transform of
the forward hadronic matrix element of two heavy--quarks
fields\footnote{Note that our definition is written in full QCD,
not in the infinite mass limit of the HQET. Another difference
with the literature is that here $F$ is a function of a
dimensionless variable, the momentum fraction $z$, rather than of
the lightcone momentum component itself. The relation with
$S(\omega)$ in \cite{Bosch:2004th} is $F(z,\mu^2)\simeq M\,
S(\bar{\Lambda}-(1-z)M)$, so $S(\omega)$ has support between $-m$
and $\bar{\Lambda}$, or in the $m\longrightarrow \infty$ limit,
between $-\infty$ and $\bar{\Lambda}$.} on the lightcone
($y^2=0$):
\begin{eqnarray}
F(z;\mu^2)=\frac{1}{4\pi}\int_{-\infty}^{\infty}
\frac{d(p_By)}{p_By}\,{\rm e}^{iz p_B y}
\left<B(p_B)\left\vert \bar{\Psi}(0)\ysl\Psi(y)\right\vert
B(p_B)\right>_{\mu^2}
\label{F},
\end{eqnarray}
and then take moments:
\begin{eqnarray}
\label{F_mom} \tilde{F}(N;\mu^2)=\int_0^1 dz
\,F(z;\mu^2)\,z^{N-1}.
\end{eqnarray}
Note that $S_N(m^2;\mu^2)$ of \eq{Soft} is nothing else but the
perturbative analogue of $F(z;\mu^2)$ which is obtained by
replacing the hadron by an on--shell heavy quark
($p^2=m^2$)~$q(p)$:
\begin{eqnarray}
F_{\PT}(z;\mu^2)=\frac{1}{4\pi}\int_{-\infty}^{\infty}
\frac{d(py)}{py}\,{\rm e}^{iz py}\left<q(p)\left\vert
\bar{\Psi}(0)\ysl\Psi(y)\right\vert q(p)\right>_{\mu^2}
\label{FPT}.
\end{eqnarray}
In moment space $\mu^2$ evolution is multiplicative.
Since the evolution factors of $\tilde{F}(N;\mu^2)$ and
$\tilde{F}_{\PT}(N;\mu^2)$ are the same, the ratio
\begin{eqnarray}
\label{FNP}
\tilde{F}_{\NP}(N)\equiv \frac{\tilde{F}(N;\mu^2)}{\tilde{F}_{\PT}(N;\mu^2)}
\end{eqnarray}
does not depend on $\mu^2$. This function incorporates power
corrections on the soft scale $m/N$ which distinguish the heavy
meson state from an on-shell heavy quark. We conclude that the
non-perturbative formula for the moments takes the form
\begin{eqnarray}
M_N=C_N(\alpha_s(m^2))\,S_N(m^2;\mu^2)\,
\tilde{F}_{\NP}(N)\,J_N(m^2;\mu^2)=M_N^{\PT}
\, \tilde{F}_{\NP}(N), \label{gen_fact_form_NP}
\end{eqnarray}
The properties of $\tilde{F}_{\NP}(N)$ will be
discussed in the following section.

\subsubsection{HQET and the OPE\label{HQET}}

The HQET (see e.g. \cite{Neubert:1993mb,Neubert:1996wg}) is a systematic way to
identify the leading contributions to QCD matrix elements in the
large mass limit. Here we would like to use this framework to
parametrize the shape function $F_{\NP}(z)$ and, in particular, to
understand its dependence on the mass definition. We recall that
in the perturbative treatment we assumed that the decaying b quark
was on-shell. Our answer for the moments of the photon--energy spectrum,
\eq{BXg_logs_N_2loop}, contains an infrared renormalon at $u=1/2$,
namely an ${\cal O}(N\Lambda/m)$ ambiguity,  which exponentiates
together with the perturbative logs. In order to trace the
cancellation of this ambiguity in the next section we first need
to understand the dependence of $F_{\NP}(z)$ on the definition of
the mass.

In the large mass limit the momentum of the heavy quark within the
hadron is naturally written as $p=mv+k$ where $v$ is the hadron
four velocity, $v\equiv p_B/M$, and $k$ is the residual momentum
which is small compared to heavy--quark mass $m$. The effective
Lagrangian describing the dynamics in the large--$m$ limit is
\begin{equation}
{\cal L}_{\eff}=h_v(x)(iv\cdot D-\delta m)h_v(x)+\cdots,
\label{L_eff}
\end{equation}
where $h_v(x)$ is the velocity--dependent rescaled quark field,
\begin{equation}
h_v(x)={\rm e}^{im_Q v\cdot x}\,\frac{1+\vsl}{2}\,\Psi(x),
\label{field_scaling}
\end{equation}
$\delta m\equiv m-m_Q$ is the residual mass term which is constant
in the large--$m$ limit, $m$ is the pole mass and the dots stand
for ${\cal O}(\Lambda/m)$ terms.

The residual mass term~\cite{Falk:1992fm} appears in the
Lagrangian because there is freedom in the choice of the mass
$m_Q$ in \eq{field_scaling}. The natural candidate for $m_Q$ is
the pole mass ($m$). However, if one is concerned with power terms
then specifying $m_Q$ as the pole mass does not justify
eliminating $\delta m$ in \eq{L_eff} because the pole mass itself
(i.e. its relation with any short--distance mass) has an infrared
renormalon at $u=1/2$ \cite{Beneke:1994sw} which implies that it
has an inherent linear ambiguity of order $\Lambda$. This
ambiguity is understood on physical grounds: owing to confinement
the full quark propagator does not have a pole. Using the pole
mass in \eq{field_scaling} would therefore render $\delta m$ zero
to any order in the coupling while non-zero and ambiguous as far
as power suppressed terms ${\cal O}(\Lambda/m)\sim \exp
\left\{-\pi/(2\alpha_s(m^2)\beta_0)\right\}$ are concerned. It was
shown in \cite{Beneke:1994sw,Neubert:1994wq} that upon including
an ambiguous $\delta m$ term in the HQET Lagrangian one can trace
the dependence of matrix elements on the mass definition and that
eventually such ambiguities cancel out in physical observables. It
is important to stress that this cancellation does not occur
within the HQET but in full QCD, and only provided that the $u=1/2$
infrared renormalon in the coefficient functions relating  HQET
matrix elements to physical observables are taken into
account.

Let us now address the {\em non-perturbative contribution} to the
b--quark distribution in the B meson, $F_{\NP}(z)$, by considering
the matrix--element definition within the HQET. In this framework
$\mu^2$ is the scale at which the effective theory is matched to
the full theory, $\Lambda^2 \ll \mu^2 \ll m^2$. Since we are not
concerned at this point with perturbative corrections we can
simply identify $F_{\NP}(z)$ with the matrix--element definition
in \eq{F} taken at Born level. Thus, in the HQET we obtain
\begin{eqnarray}
\label{F_HQET} F_{\NP}(z)=\int_{-\infty}^{\infty} \frac{d(v \cdot
y)}{4\pi}\,{\rm e}^{iM(v\cdot y)
\left(z-\frac{m_Q}{M}\right)}\left<B(Mv)\left\vert \bar{h}_v(0)\,
P{\rm e}^{-i\int_0^1ds y_{-}A_{+}(sy)}\,h_v(y)\right\vert B(Mv)\right>,\nonumber \\
\end{eqnarray}
where we used \eq{field_scaling} and substituted $p_B=Mv$. We also
wrote explicitly the path--ordered exponential making the matrix
element gauge invariant. Note that the normalization of the states
in \eq{F_HQET} is
\[\left< B(Mv)\left\vert \bar{h}_v(0)h_v(0)\right\vert B(Mv)\right>=2M.\]

Next we would like to expand the {\em non-local} lightcone
operator in \eq{F_HQET} in terms of {\em local} operators. The
matrix elements of these operators can be expressed by
the parameters of the HQET. A similar expansion was performed in
the past, see e.g.~\cite{JR,Bigi:1993ex,Mannel:1994pm}. At
difference with previous literature, however, we do not put
$\delta m$ to zero because we want to distinguish between matrix
elements that depend on $\delta m$ and these that do not. The
expansion yields:
\begin{equation}
 \hspace*{-6pt}
\frac{1}{2M} {\left<B(Mv)\left\vert \bar{h}_v(0)
\,P{\rm e}^{-i\int_0^1ds y_{-}A_{+}(sy)}\,h_v(y)\right\vert B(Mv)\right>}=
1+\sum_{k=1}^{\infty}\frac{a_k}{k!}  \,(-iv\cdot y)^k,
\label{local_expansion}
\end{equation}
where
\begin{equation}
(v_{+})^k\,
a_k\equiv \frac{1}{2M}
{\left<B(Mv)\left\vert \bar{h}_v(0)(iD_{+})^k \,h_v(0)\right\vert B(Mv)\right>}.
\label{HQET_mat_elem}
\end{equation}
The matrix elements $a_k$ depend on $\delta m$, which is a
parameter of the effective Lagrangian of \eq{L_eff}. Since $\delta
m$ represents the ambiguity that should cancel against the $u=1/2$
renormalon in the perturbative coefficient function we would like
to isolate this dependence.

Using the equation of motion \[i(v\cdot D)h_v(x)=\delta m h_v(x)\]
we can immediately fix the first parameter in \eq{local_expansion}, $a_1=\delta m$.
Using further the matrix element corresponding to the kinetic energy
of the b quark inside the meson (see e.g.~\cite{Bigi:1993ex,Neubert:1993um}) we
 can fix also the second coefficient $a_2$.
Let us define
\begin{equation}
\frac{1}{2M}{\left<B(Mv)\left\vert \bar{h}_v(0)
(g_{\mu\nu}-v_{\mu}v_{\nu})
iD^{\mu}iD^{\nu} \,h_v(0)\right\vert
B(Mv)\right>}=\lambda_1.
\end{equation}
As discussed in \cite{Neubert:1993um} $\lambda_1<0$. It is
straightforward to check by a change of variables in the
functional integral,
\begin{eqnarray}
\delta m &\longrightarrow & \widetilde{\delta m}, \nonumber \\
h(y)&\longrightarrow &  \widetilde{h}(y)= {\rm e}^{i(\widetilde{\delta m}-\delta m)\, vy}\,h(y),
\end{eqnarray}
that, like the
Lagrangian, $\lambda_1$ is independent of $\delta m$. Having
established this we express $a_2$ in terms of $\lambda_1$ exposing
explicitly the dependence on $\delta m$: $a_2=(\delta
m)^2-\lambda_1/3$. Similarly, to deal with $a_3$ we use the
definition
\begin{eqnarray}
\frac{1}{2M}{\left<B(Mv)\left\vert \bar{h}_v(0) v_{\nu}D_{\mu}F^{\mu\nu}
\,h_v(0)\right\vert B(Mv)\right>}&=&-3f_3.
\end{eqnarray}
Also $f_3$ is independent of $\delta m$.
We can now express the third coefficient as $a_3=(\delta
m)^3-\lambda_1\delta m+f_3$.
It is convenient to define $f_2\equiv
-\lambda_1/3>0$ and then the first three coefficients are:
\begin{eqnarray}
a_1&=&\delta m,\nonumber \\
a_2&=&(\delta m)^2+f_2,\nonumber\\
a_3&=&(\delta m)^3+3f_2\delta m+f_3.
\end{eqnarray}
This procedure can be continued to higher orders. At each order $n$
the coefficient $a_n$ contains at least one new parameter which is
unrelated to the definition of the mass and thus enters with
$(\delta m)^0$; in addition $a_n$ contains a polynomial of order
$n$ in $\delta m$. We observe, however, that the $\delta
m$--dependent terms have a very specific structure (to any order):
they exponentiate. The r.h.s. of \eq{local_expansion} can be
resummed, getting
\begin{eqnarray}
\label{exp_of_delta_m}
\frac{1}{2M} {\left<B(Mv)\left\vert \bar{h}_v(0)
\,P{\rm e}^{-i\int_0^1ds y_{-}A_{+}(sy)}\,h_v(y)\right\vert B(Mv)\right>}&=&
\\
{\rm e}^{-iv\cdot y \delta m}
\,\times\, \left[1+(-iv\cdot y)^2\frac{f_2}{2}
+(-iv\cdot y)^3\frac{f_3}{6} +\cdots\right]\nonumber
&\equiv &
{\rm e}^{-iv\cdot y \delta m}\,{\cal F}(-iv\cdot y\Lambda),
\end{eqnarray}
where the sum in the square brackets, and therefore the function
${\cal F}(-iv\cdot y\Lambda)$, are entirely free of the $u=1/2$
renormalon ambiguity; $\Lambda$ appearing in ${\cal F}(-iv\cdot
y\Lambda)$ is the QCD scale: it {\em must not} be identified with
the ambiguous $\bar{\Lambda}=M-m$. The exponential factor in \eq{exp_of_delta_m}
can be understood from general considerations: it is a reminiscent
of the scaling of the fields in \eq{field_scaling}. The ambiguity
in the mass used to {\em define} the HQET reappears as an
exponential factor in the parametrization of the matrix element of
\eq{local_expansion}.

Using \eq{exp_of_delta_m}, \eq{F_HQET} becomes
\begin{eqnarray}
\label{F_HQET_exp} F_{\NP}(z)=\int_{-\infty}^{\infty}
\frac{d\omega}{2\pi}\,{\rm e}^{i\omega\left(z-\frac{m_Q+\delta
m}{M}\right)} \,{\cal F}(-i\omega\Lambda/M)+{\cal O}(\Lambda/M),
\end{eqnarray}
exhibiting the fact \cite{Falk:1992fm} that the residual mass term
$\delta m$ enters matrix elements only in the combination
$m_Q+\delta m=m$.

Next, using \eq{F_HQET_exp} together with the definition of the
moments, \eq{F_mom}, one can show that to leading order in $1/N$,
\begin{eqnarray}
\label{F_NP_result}
\tilde{F}_{\NP}(N)&=&{\rm
e}^{-(N-1)\bar{\Lambda}/M}\, {\cal F}((N-1)\Lambda/M)+ {\cal
O}(\Lambda/M)
\\\nonumber &=& {\rm e}^{-(N-1)\bar{\Lambda}/M}\,
\bigg[1+\frac{f_2}{2!}\left(\frac{(N-1)\Lambda}{M}\right)^2+\frac{f_3}{3!}
\left(\frac{(N-1)\Lambda}{M}\right)^3\,+\,\cdots\bigg] \,+ {\cal
O}(\Lambda/M),
\end{eqnarray}
where we used the definition $\bar{\Lambda}=M-m=M-m_Q-\delta m$ in
the exponential factor. To derive this relation one first converts
the Mellin integral over $z$ into Laplace, neglecting ${\cal
O}(1/N)$ (or, equivalently, ${\cal O}(\Lambda/M)$) corrections.
The $z$ integral is then straightforward. Finally, the $\omega$
integral can be done by closing a contour in the complex plane
picking up the residue at $\omega=i(N-1)$. In the following, as in
\eq{F_NP_result}, we shall explicitly use $m$ (and
$\bar{\Lambda}$) rather than $m_Q$. The reader should keep in mind
that this parameter is ambiguous.

Note that \eq{F_NP_result} is scaling law. For example,
it excludes corrections scaling like ${\cal O} (N^2\Lambda/M)$.
Put differently, it implies that there is one natural way to take
the large $N$ and large $M$ limits, namely with the ratio
$N\Lambda/M$ fixed. 

Note also that in \eq{F_NP_result} we have
chosen to write everywhere $N-1$ rather than $N$ in order to impose the
requirement that $\tilde{F}_{\NP}(N=1)=1$ (which is fixed by the
overall normalization of the moments, $M_{N=1}\equiv 1$) without involving
additional ${\cal O} (1/N)$ terms. This way our approximation for
$\tilde{F}_{\NP}(N)$ captures correctly not only the large--$N$ 
limit but also the $N=1$ limit.

We stress again the distinction between the first factor in
\eq{F_NP_result}, ${\rm e}^{-(N-1)\bar{\Lambda}/M}$ which contains
a mass--related renormalon ambiguity at $u=1/2$ and the second
factor, ${\cal F}((N-1)\Lambda/M)$, which is free of this
ambiguity. ${\cal F}((N-1)\Lambda/M)$ does not depend on the quark
mass but only on the physical meson mass through the ratio
$(N-1)\Lambda/M$. Besides this fact, the most important property
of this function is that its linear term in the small
$(N-1)\Lambda/M$ expansion vanishes, as shown in the second line
of~\eq{F_NP_result}.

Having established that ${\cal F}((N-1)\Lambda/M)$ itself is free
of any $u=1/2$ renormalon ambiguity, it is important to recall
that this function, or alternatively the values of matrix elements
$f_n$, do contain (in general) other ($u\neq 1/2$) renormalon
ambiguities of {\em ultraviolet} origin. Similarly to
higher--twist corrections to deep inelastic structure functions
these matrix elements need to be defined by specifying a
regularization prescription for the renormalons which corresponds
to the one used for infrared renormalons in the perturbative
coefficient function.

At this point it is useful to compare \eq{F_NP_result} with the
renormalon--based ansatz of \eq{BXg_powers}. The leading term,
${\cal O}((N-1)\Lambda/M)$ exponentiates in both. Note, however,
that the reasoning that lead us to deduce that exponentiation
occurs was completely different: in \eq{BXg_powers} it was based on
the exponentiation of perturbative soft radiation while in
\eq{BXg_powers} it was based on the dependence of the heavy--quark
field on the mass. Higher--order terms in the HQET analysis do not
exponentiate but this cannot be considered as contradiction
between the two, as we do not know how to sum the series of power
terms. A more significant difference is the presence of a second
power ${\cal O}(((N-1)\Lambda/M)^2)$ in \eq{F_NP_result},
associated with the kinetic energy of the heavy quark inside the
meson in contrast with its absence in \eq{BXg_powers}. We recall
that the latter is a consequence of the vanishing of  the
corresponding renormalon residue in the perturbative result,
\eq{BXg_logs_N_2loop}, in the large--$\beta_0$ limit\footnote{The
vanishing of the renormalon associated with the kinetic energy in
the single--dressed--gluon approximation has been noted and
analyzed in detail in the past, see e.g.
\cite{Martinelli:1995zw}.}.  In spite of the vanishing of this
renormalon, it is sensible to allow an ${\cal
O}(((N-1)\Lambda/M)^2)$ term in a phenomenological model for power
corrections. To understand this term from the renormalon
perspective note that: (1) Generally, absence of a renormalon does
not imply the vanishing of the corresponding non-perturbative
parameter, only the vanishing of its ambiguity. Thus,
\eq{BXg_powers} should only be considered as a minimal
parametrization, not a constraining one. (2) It has been shown
in~\cite{Neubert:1996zy} that in the case of the kinetic energy of
the b quark a renormalon singularity does appear when going beyond
the single--dressed--gluon approximation.

\subsubsection{Cancellation of the $u=1/2$ renormalon\label{cancellation}}

It is well known \cite{BBZ,Neubert:1994wq} that the total
semi-leptonic decay width,
\begin{eqnarray}
\Gamma(\bar B\longrightarrow X_u l \bar{\nu}) =\frac{G_F^2\left\vert
V_{ub}\right\vert^2m^5}{192\pi^3}\left[1-\frac{2\alpha_s(m^2)}{3\pi}
\left(\pi^2-\frac{25}{4}\right)+\cdots \right],
\end{eqnarray}
is free of any $u=1/2$ renormalon ambiguity. If the pole mass is
used in the calculation an intricate cancellation occurs between
the $u=1/2$ ambiguity contained in the overall factor~$m^5$ and
that associated with the sum of the perturbative series in the
square brackets. If a short--distance mass is used instead, these
two factors are separately free of such ambiguities. Here we 
show that similar cancellation takes place in differential 
decay spectra.

To demonstrate the cancellation of ambiguities we shall resum the
perturbative series in the single--dressed--gluon (large--$\beta_0$) approximation. 
One difference with respect to the
total width is that when considering the spectrum we need to take
into account exponentiation: Sudakov exponentiation on the
perturbative side and the exponentiation of the $\delta m$ terms
on the non-perturbative side.

Let us first consider the ambiguity in the non-perturbative
function in \eq{F_NP_result} in this approximation. The result is:
\begin{eqnarray}
\label{F_NP_PV} \tilde{F}_{\NP}(N)&\simeq&{\rm
e}^{-(N-1)({M-m})/{M}}\, {\cal F}((N-1)\Lambda/M) \\
\nonumber&=& {\rm e}^{(N-1)\Delta m/M } \,\times\, {\rm
e}^{-(N-1)(M-m_{\PV})/M} {\cal F}((N-1)\Lambda/M)
\end{eqnarray}
with
\begin{equation}
\frac{\Delta m}{m}\equiv \frac{m-m_{\PV}}{m}\simeq
\frac{C_F}{2\beta_0}\left[\int_0^{\infty}du \,T(u)
\left(\frac{\Lambda^2}{m^2}\right)^u \,\frac{{\rm e}^{c
u}}{\frac12-u}\,-\,{\rm PV}\int_0^{\infty}du \,T(u)
\left(\frac{\Lambda^2}{m^2}\right)^u \,\frac{{\rm e}^{c
u}}{\frac12-u}\right], \label{Delta_m}
\end{equation}
where, as before, we define $\Lambda$ in ${\overline{\rm MS}}$ so
$c=5/3$. Here $m_{\PV}$ stands for the pole mass in which the
$u=1/2$ renormalon is regularized by the principal--value
prescription. \eq{Delta_m} is correct to leading order in
$1/\beta_0$; we also neglected higher power corrections ${\cal
O}(\Lambda^3/m^3)$. The explicit expression for $\frac{\Delta
m}{m}$ in the large--$\beta_0$ limit was obtained based on the
known relation \cite{BBZ,Beneke:1994sw,Neubert:1994wq} in this
limit between the pole mass and short--distance mass definitions.
In particular the relation with the ${\overline{\rm MS}}$ mass is
\begin{equation}
{\frac{m-m_{\MSbar}}{m}}= \frac{C_F}{2\beta_0}
\int_0^{\infty}du\,T(u) \left(\frac{\Lambda^2}{m}\right)^u\,
\left[{\rm
e}^{cu}\frac{3(1-u)\Gamma(u)\Gamma(1-2u)}{\Gamma(3-u)}+\frac{1}{2u}\,\widetilde{G}(u)\right],
\label{Delta_m_MSbar}
\end{equation}
where $\widetilde{G}(u)$ depends on the ${\overline {\rm MS}}$
counter terms for the mass\footnote{In ${\overline {\rm MS}}$
$\widetilde{G}(u)$ is given by $\sum_{n=0}^{\infty}g_nu^n/n!$ with
$g_n$ defined as the expansion coefficients of
\[
G(u)=-\frac13
\frac{(3+2u)\Gamma(4+2u)}{\Gamma(1-u)\Gamma(2+u)^2\Gamma(3+u)}.
\]
} and it is free of renormalons. Note that since the difference
between masses is ${\cal O}(1/\beta_0)$ it does not matter (to
leading order in $1/\beta_0$) which mass is used to normalize the
mass difference in the equations above.

Let us turn now to the perturbative side. We recall that owing to
the integration over $x$ near $1$, the perturbative moments
$M_N^{\PT}$ of \eq{BXg_logs_N_2loop} have a renormalon ambiguity
at $u=1/2$ which is associated with the soft scale $m/N$. Using in
\eq{BXg_logs_N_2loop} the principal--value prescription to define
$\left.M_N^{\PT}\right\vert_{\PV}$ we obtain:
\begin{eqnarray}
\label{MN_PT_PV}
 M_N^{\PT}\simeq \left.M_N^{\PT}\right\vert_{\PV}
\,\times\, \exp \Bigg\{
\frac{C_F}{2\beta_0}\bigg[&&\!\!\!\!\int_0^{\infty}du \,T(u)\,
\left(\frac{\Lambda^2}{m^2}\right)^u\,\frac{{\rm
e}^{cu}}{\frac12-u}\left(N^{2u}-1\right)\big(1_{\rm I}-2_{\rm
II}+0_{\rm
III}\big)\nonumber \\
-{\rm PV}&&\!\!\!\!\int_0^{\infty}du \,T(u)\,
\left(\frac{\Lambda^2}{m^2}\right)^u\,\frac{{\rm
e}^{cu}}{\frac12-u}\left(N^{2u}-1\right)\big(1_{\rm I}-2_{\rm
II}+0_{\rm III}\big)
 \bigg] \Bigg\},\nonumber \\
\end{eqnarray}
where the subscripts I, II and III indicate the diagram in Fig.
\ref{fig:diagrams} from which each contribution originates in the
Feynman gauge\footnote{A similar analysis for the contributions of
different diagrams to the renormalon residue in the total decay
rate was done in \cite{BBZ}. There an additional diagram,
the self energy of the heavy line, contributes.}, {\em cf.} \eq{B_BXSG}. Note that in
\eq{MN_PT_PV} we have isolated the $u=1/2$ pole ignoring subleading
power corrections which are unrelated to this renormalon.
We stress that although the
large--$N$ limit has been taken to arrive at
\eq{BXg_logs_N_2loop}, our result for the $u=1/2$ residue (in the
large--$\beta_0$ limit) as expressed by \eq{MN_PT_PV} is exact: as
one can verify by taking moments of the exact expression in $x$
space, \eq{B_BXSG}, there are no subleading $1/N$ corrections
which have renormalon ambiguities at $u=1/2$. The reason is that
this renormalon is directly related to the integration over $x$ in
the singular $x=1$ limit.

Finally, substituting in~\eq{gen_fact_form_NP} the perturbative
and non-perturbative factors of \eq{MN_PT_PV} and \eq{F_NP_PV},
respectively, we get
\begin{eqnarray}
\label{MN_cancel} M_N=M_N^{\PT}\tilde{F}_{\NP}(N) = \left.
M_N^{\PT}\right\vert_{\PV}\, {\rm e}^{-(N-1)(M-m_{\PV})/M} {\cal
F}((N-1)\Lambda/M),
\end{eqnarray}
where the ambiguous exponential factors cancel out in the product.
The result is entirely free of the $u=1/2$ renormalon ambiguity.
Moreover, it is {\em independent} of the regularization
prescription for the $u=1/2$ renormalon pole; what is important is
that the {\em same} prescription (here principal value of the Borel sum) is used on both the
perturbative Sudakov exponent in $\left.
M_N^{\PT}\right\vert_{\PV}$ and on the pole mass in ${\rm
e}^{-(N-1)(M-m_{\PV})/M}$. Let us stress that \eq{MN_cancel}
resums only power corrections depending on the soft scale, $M/N$:
subleading power corrections such as ${\cal O}(\Lambda/M)$ or
${\cal O}(\Lambda^2N/M^2)$ are neglected here.

The implications of \eq{MN_cancel} become more intuitive upon returning
to $x$ space. Using the inverse Mellin transform\footnote{The
contour ${\cal C}$ goes from $k-i\infty$ to $k+i\infty$ to the
right of the singularities of the integrand.} we have
\begin{eqnarray}
\label{back_to_x}
 \frac{1}{\Gamma_{\tot}}\frac{d\Gamma(x)}{dx}&=&
\frac{1}{2\pi i}\int_{\cal C} dN \,x^{-N} M_N\nonumber \\&\simeq&
\frac{1}{2\pi i}\int_{\cal C} dN \,{\rm e}^{(N-1)\left(1-x-\frac
{M-m_{\PV}}{M}\right) }\, \left. M_N^{\PT}\right\vert_{\PV}\,{\cal
F}((N-1)\Lambda/M),
\end{eqnarray}
where in the second line we neglected ${\cal O} (1/N)$ terms. If
we neglect the effect of ${\cal F}((N-1)\Lambda/M)$ in
\eq{back_to_x} altogether, the only effect of confinement is to
shift the perturbative spectrum in $x$ by the energy fraction of
the light degrees of freedom in the meson, $(M-m)/M$, away from
the endpoint, towards smaller values of $x$:
\begin{equation}
\frac{1}{\Gamma_{\tot}}\frac{d\Gamma(x)}{dx}\sim
\frac{1}{\Gamma_{\tot}}\frac{d\Gamma^{\PT}_{\PV}(x+\frac
{M-m_{\PV}}{M})}{dx}.
\label{Decay_shift}
\end{equation}
Note that the {\em shifted} spectrum does not depend anymore on
the quark mass definition or on the fact that the principal--value
prescription was used (for both the Sudakov exponent and the pole
mass). The finding that the leading effect of confinement is a
shift of the perturbative spectrum is not surprising nor unique to
$B$ physics. It is a general feature of differential cross
sections near kinematic thresholds. For example, a shift occurs in
event--shape distributions which peak near the two jet
limit~\cite{Korchemsky:1995zm,Dokshitzer:1997ew,Gardi:2001ny}.
What is special though is that in the $B$--decay case one can fix
the {\em magnitude} of the shift in terms of a familiar
non-perturbative parameter.

It should be stressed that this shift has nothing to do with the
fact that the physical endpoint is at $E_\gamma=M/2$ whereas the
perturbative one is at $E_\gamma=m/2$; note that \eq{Decay_shift}
concerns the spectrum in the dimensionless scaling variable $x$.
In fact, when addressing the close vicinity of endpoint the
function ${\cal F}((N-1)\Lambda/M)$ cannot be neglected in
\eq{back_to_x}. Physically, ${d\Gamma(x)}/{dx}$ has support
for $0< x< 1$. The perturbative spectrum which is calculated using
the principal--value prescription in moment space does not respect
the physical support properties and these should be recovered in
\eq{back_to_x} when the shift generated by the exponential as well
as the additional smearing generated by the subleading power
corrections in ${\cal F}((N-1)\Lambda/M)$ are taken into
account. If $(N-1)\Lambda/M$ is not too large the leading power
correction in ${\cal F}((N-1)\Lambda/M)$ is that of the kinetic
energy of the $b$ quark inside the meson, $f_2$. However, near the
endpoint higher power terms become as important.

The reader should also keep in mind that the spectrum computed
by~\eq{back_to_x} is a priori just a rough description of the
physical one at $x$ values away from the endpoint region: there
are additional ${\cal O}(1/N)$, or ${\cal O}(\Lambda/M)$, effects
which were neglected altogether. Formally, our analysis
(perturbative and non-perturbative) applies only at large $N$.
Having fixed, in addition to the large--$N$ limit, the $N=1$ moment
one can hope that the numerically most important corrections 
in the first few moments are accounted for, but this is not guaranteed
and it will be eventually decided by the data. Experience in other
applications \cite{Gardi:2001ny,CG} is very encouraging in this respect.

\section{Semi-leptonic decay\label{SL_decay}}

In this section we consider the semi-leptonic decay
$\bar B\longrightarrow X_u l^-\bar{\nu}_l$ and perform a DGE
calculation in analogy with the radiative decay presented
above. Let us start by describing the kinematics. We call the
$b$--quark momentum $p$, and the momentum of the leptons
$k_{\bar{\nu}_l}$ and $k_l$. Let us also denote the total momentum
of the lepton system by $q\equiv k_{\bar{\nu}_l}+k_l$. We will
express the distribution in terms of $x\equiv 2pk_{l}/m^2$,
$y\equiv 2pq/m^2$ and the leptonic invariant mass fraction
$z\equiv q^2/m^2$. In the $B$ rest frame $y$ has the meaning of
the total leptonic energy ($E$) fraction, $y= 2E/m$, while $x$ is
the electron (or muon) energy fraction. The hadronic system (a
$u$--quark jet) has momentum $l$ and invariant mass
\begin{equation}
l^2=q^2+m^2(1-y)=m^2(1-y+z).
\end{equation}

The semi-leptonic decay rate is given by
\begin{equation}
\Gamma=\frac{G_F^2\vert v_{ub}\vert^2}{2m} \,\int d{\rm LIPS}(p\longrightarrow
k_{\bar{\nu}_l}+k_l+l)\,\,{\rm Im} \left\{
L^{\mu\nu}T_{\mu\nu}\right\}.
\end{equation}
Writing the phase--space integral in terms of $x$, $y$ and $z$ we have:
\begin{eqnarray}
 \label{sl_ps}
\!\!\Gamma\!=\! \frac{G_F^2\vert v_{ub}\vert^2\,m^3}{64\pi^4}
\,\int_0^1\!dz\, \int_z^1 dx \int_{ x+\frac{z}{x}}^{1+z}\! \!dy
\,{\rm Im}\left\{A_{\rm Born}+\frac{C_F\alpha_s}{4\pi}
\left[A_{\rm I}+2A_{\rm II}+A_{\rm III}\right]\right\},
\end{eqnarray}
where $A_{\rm Born}$ stands for the Born diagram contracted with
the leptonic tensor, and $A_{\rm I,II,III}$ correspond to the
three one--loop diagrams in the Feynman gauge, contracted with the
leptonic tensor. The Born level result is: $A_{\rm Born}=
-{\displaystyle  {4\,m^{4}\,(x-z)\,(y-x)}/{{l^2}}}$. Taking the
imaginary part we get
\begin{eqnarray}
\label{SL_Born}
{\rm Im}\left\{A_{\rm Born}\right\}&=&4m^2(x-z)\,
(y-x)\times \pi\delta(1-y+z).
\end{eqnarray}

Next, we compute the one--loop diagrams with a dressed gluon. The
calculation is more complicated than that of the radiative decay
owing to the presence of additional kinematic variables. In
Appendix \ref{sl_calc} we summarize the full result as a double
Feynman parameter integral. The interest here, however, is in the
specific kinematic limit where Sudakov logs appear: based on the
Born--level calculation this limit is $y\longrightarrow 1+z$.

We observe that the phase--space structure of
\eq{sl_ps} implies that the region $y\simeq 1+z$ becomes relevant
for $x\longrightarrow 1$: the lower limit in the integration over~$y$ 
approaches $1+z$, which is also the upper limit. Thus, if one
considers the spectrum near the endpoint $x\longrightarrow 1$ the
Sudakov region is indirectly selected. Nevertheless, one can
approach the Sudakov limit differently, e.g. by considering
directly the region where the hadronic system has small invariant
mass. Below we first present a triple differential spectrum to
leading power in $(1+z-y)$ allowing one to approach the Sudakov
limit in different ways. Then we focus on the specific example of
the single differential rate in $x$ to leading power in $(1-x)$.

In Appendix \ref{JETSOFT} we compute all the singular terms of the
semi-leptonic decay in
the limit $y\longrightarrow 1+z$  in
the single--dressed--gluon approximation. Combining the jet and soft
functions, \eq{Im_T_total_jet} and \eq{Im_T_total_soft},
respectively, and using \eq{sl_ps} we arrive at the following
triple differential decay rate to leading power in $(1+z-y)$:
\begin{eqnarray}
\label{sl_logs_fact}
&&\hspace*{-10pt}\frac{1}{\Gamma_{\tot}}\frac {d\Gamma(y,x,z)}{dy dx dz}=12\,(x-z)(1+z-x)\,
\left\{\delta(1+z-y)+\frac{C_F}{2\beta_0}\int_0^{\infty}du\,
T(u)\left(\frac{\Lambda^2}{m^2}\right)^u {\rm e}^{cu}\right.
\nonumber \\
&&\hspace*{-10pt} \left.
\,\,\times\left[\frac{2}{u}\frac{1-u}{1-z}\Big(\frac{1+z-y}{1-z}\Big)^{-1-2u}
-\left(\frac{2}{u}+\frac{1}{1-u}+\frac{1}{2-u}\right)
\frac{\sin \pi u}{\pi u}\Big(1+z-y\Big)^{-1-u}
\right]\right\},
\end{eqnarray}
where we normalized the spectrum dividing by the total decay
width:
\begin{equation}
\Gamma_{\tot}=\Gamma_0\times(1+{\cal O}(\alpha_s)), \quad\quad
\Gamma_0=\frac{G_F^2\vert v_{ub}\vert^2 m^5}{192\pi^3},
\end{equation}
where $\Gamma_0$ is the Born--level width. \eq{sl_logs_fact}
is the analogue of \eq{BXg_logs_fact}.

We note that the general $u$-dependent structure of the soft and
the jet functions in \eq{Im_T_total_soft} and \eq{Im_T_total_jet}
is the same as in the radiative decay ({\em cf.}
\eq{BXg_logs_fact}), the difference being restricted to the $z$
dependence of the two scales involved. Here the soft function
depends on $m(1+z-y)/(1-z)$ and the jet function on $m^2(1+z-y)$.
Clearly, in the limit $z\longrightarrow 0$ \eq{sl_logs_fact}
reduces to \eq{BXg_logs_fact}, where $y$ in the former takes the
role of $x$ in the latter.

We recall that these functions appear not only in heavy--flavor
decay spectra, but also in other differential cross sections near
a kinematic threshold. The perturbative soft function is
associated with {\em radiation off a heavy quark close to its mass
shell}, and it therefore describes also the heavy--quark
fragmentation function \cite{CG}. The jet function~\cite{Sterman:1986aj,CSS,CT,Catani:1992ua,Korchemsky:1992xv,CLS,CMW,Akhoury:1998gs}
 occurs even more frequently: it is not specific to heavy--quark physics. It
describes {\em radiation off an unresolved jet} with a constrained
invariant mass $l^2$. In the framework of DGE the same expression for the 
jet function has already 
been obtained in the context of deep inelastic structure functions at large
Bjorken $x$~\cite{Gardi:2001di,Gardi:2002bk,Gardi:2002xm}, single--particle 
inclusive cross sections in $e^+e^-$ annihilation for light~\cite{Gardi:2001di} 
and heavy quarks~\cite{CG} and event--shape distributions~\cite{Gardi:2001ny,Gardi:2003iv}.

Infrared and collinear safety requires cancellation of $u=0$ poles
between the soft and the jet functions. Indeed, the result is
finite: the expansion of the Borel function in the square brackets
of \eq{sl_logs_fact} is
\begin{equation}
\frac{-1}{1+z-y}\left(2\ln(1+z-y)-4\ln(1-z)+\frac72 \right)+{\cal
O}(u).
\end{equation}
This (times $C_F/(2\pi)$) is the logarithmically--enhanced part 
of the coefficient of $\alpha_s$ in the
perturbative expansion. One immediately identifies the double--
and the single--logarithmic terms of $(1+z-y)$. At difference with
the radiative--decay case, there are also single logs of $(1-z)$
originating from the soft function, which mix with those of
$(1+z-y)$. Note that $z\longrightarrow 1$ corresponds to another
singular limit (not considered here) where the invariant mass of
the lepton pair approaches the total b--quark mass.

To resum soft--gluon radiation, \eq{sl_logs_fact} needs to be
exponentiated in moment space in analogy with
\eq{BXg_logs_N_2loop}. Natural moment--space definitions are with
respect to $y-z$, or the electron energy fraction $x$. In the
following we concentrate on the latter possibility, which is
useful for describing the single differential rate with respect to
$x$. To this end we need to integrate the triple differential
rate, \eq{sl_logs_fact}, over the phase space: for any fixed~$x$
($0\leq x\leq 1$) the available phase space ({\em cf.} \eq{sl_ps})
is
\begin{eqnarray}
\label{app_phase_space} \int_0^x dz\int_{x+\frac{z}{x}}^{1+z}
dy\,&=&\, \int_1^{1+x}dy
\int_{y-1}^{x(y-x)}dz+\int_x^1dy\int_0^{x(y-x)}dz\nonumber \\&\simeq&
\int_1^{1+x}dy (2-y)\int_{\frac{1+x^2-xy}{2-y}}^1d\xi\simeq
\int_1^2dy(2-y)\int_x^1d\xi,
\end{eqnarray}
where in the first step we changed the order of integration
keeping the exact phase space; in the second we neglected the
second term which is ${\cal O}(1-x)$ with respect to the first; in
the third we changed variables to $\xi\equiv ({1-z})/({2-y})$; and
in the fourth approximated the integration limits neglecting
${\cal O}(1-x)$ terms.

Integrating the single--dressed--gluon result,~\eq{sl_logs_fact},
over this approximate phase space we retain exactly the leading power
in $(1-x)$. In order to preform exponentiation we now take
moments with respect to $x$,
\begin{equation}
M_N^{\rm sl}=\int_0^1 dx \,x^{N-1}\,
\int_1^2dy(2-y)\int_x^1d\xi\,\, \frac{1}{\Gamma_{\tot}}
\left.\frac {d\Gamma(y,x,z)}{dy dx dz}\right\vert_{z=1-\xi(2-y)}.
\label{sl_moments}
\end{equation}
Using~\eq{sl_logs_fact}, the DGE formula takes the form:
\begin{eqnarray}
\label{SL_logs} &&M_N^{\rm sl}=C_N^{\rm sl} (\alpha_s(m^2))\,
\exp \bigg\{
\frac{C_F}{2\beta_0}\int_0^{\infty}du \,T(u)\,
\left(\frac{\Lambda^2}{m^2}\right)^u\,{\rm e}^{cu}\,
 \bigg[ \frac 2u
(1-u)\,\Gamma(-2u)\left(N^{2u}-1\right)
 \nonumber \\
 &&\hspace*{30pt} -
\left(\frac2u+\frac{1}{1-u}+\frac{1}{2-u}\right) \frac{6}{(2-u)(3-u)}\,\frac{\sin\pi
u}{\pi u}\,\Gamma(-u)\left(N^{u}-1\right)\bigg] \bigg\},
\end{eqnarray}
where we absorbed an overall factor ($\sim 2/N$ at large $N$) into
the coefficient $C_N^{\rm sl}$ and exponentiated the logarithmic
terms. There are two differences with respect to the radiative
decay case, \eq{BXg_logs_N_2loop}. The most important one that
here $C_N^{\rm sl}$  is ${\cal O} (1/N)$ rather than ${\cal
O}(1)$. The other is the factor ${6}/[{(2-u)(3-u)}]$ in the second
term which emerges from a weighted integral over the jet function,
the integral over $y$ in \eq{sl_moments}.

Upon expanding the exponential and the Borel sum to order
$\alpha_s$ we obtain:
\begin{eqnarray}
M_N^{\rm sl}=C_N^{\rm sl} (\alpha_s(m^2))\left\{1+\frac{C_F\alpha_s}{2\pi}\left[-\ln^2N
+\left(\frac{31}{6}-2\gamma_E\right)\ln N\right]+{\cal O}(\alpha_s^2)
\right\}.
\end{eqnarray}
Note that the constant ($N$--independent) terms are all in $C_N^{\rm
sl} (\alpha_s(m^2))$. This result for the leading and
next--to--leading logarithmic terms agrees with the known ${\cal
O}(\alpha_s)$ coefficient, see
\cite{Jezabek:1988ja,DeFazio:1999sv}.

It is now straightforward to
match our resummed spectrum, \eq{SL_logs}, to the exact ${\cal O}(\alpha_s)$
result to account for subleading powers of $1/N$.
The matching coefficient function is given by:
\begin{eqnarray}
\label{C_sl}
&&C_N^{\rm sl} (\alpha_s(m^2))\,=\,  \frac {2\,(N + 5)}{(N + 2)\,(N + 3)}\,\times\Bigg\{1+
\frac{C_F\alpha_s}{2\pi}\bigg\{\bigg[
 - \left(\Psi (N) + \gamma_E \right)^{2}+\Psi_1(N)- { \frac {\pi ^{2}}{6}}
\nonumber \\
&&
 + \left( - { \frac {5}{N + 1}
}  + { \frac {31}{6}}  - { \frac {6}{5
\,(N + 5)}}  + {\displaystyle \frac {31}{5\,N}} \right)\,\left(\Psi (N) +
\gamma_E \right)   + {
\frac {11}{3\,(N + 2)}}  - { \frac {11}{150\,N}}
 \nonumber\\
&&\mbox{} - { \frac {15}{2}}  + { \frac {
211}{150\,(N + 5)}}  - { \frac {5}{(N + 1)^{2}}}  +
{ \frac {4}{3\,(N + 3)}}  + { \frac {31
}{5\,N^{2}}}  + { \frac {5}{3\,(N + 1)}}  \bigg]
\nonumber \\
&&-\left[ { - \mathrm{ln}^{2} N +
\left({\frac {31}{6}- 2\,\gamma_E}  \right)\,\mathrm{ln}N}\right]\bigg\}
+{\cal O}\left(\alpha_s^2\right)\Bigg\}.
\end{eqnarray}

\eq{SL_logs} together with \eq{C_sl} summarizes our perturbative
result for the electron spectrum in the semi-leptonic decay. As
usual, the perturbative sum in \eq{SL_logs} needs a prescription
for renormalon ambiguities. Once regularized, this spectrum can be
combined with non-perturbative corrections and then converted to
$x$ space according to \eq{back_to_x}. It is natural to use the
same regularization prescription as in the radiative decay (the
principal value). Only then can we identify the shape function
between the two physical processes.

\section{Conclusions\label{conc}}

We presented here a new approach to describe inclusive B--decay
spectra in QCD focusing on the endpoint region, $x\longrightarrow
1$.  We observe that precise description of the spectra near the endpoints 
requires power--like separation between perturbative and non-perturbative 
contributions on the soft scale, $M/N$. Our
approach to the problem is based on Dressed Gluon Exponentiation
(DGE) and it differs from conventional Sudakov resummation by
running--coupling effects, renormalons, which probe the inherent
power--like ambiguity of the Sudakov exponent. The 
non-perturbative contribution to the b-quark distribution in the meson 
is, by itself, ambiguous, so it can only be defined in correspondence with 
the regularization of the renormalons in the perturbative result. For the leading
renormalon, ${\cal O}(N\Lambda/M)$, we have explicitly shown how a
regularization--prescription (or mass--scheme) independent,
well-defined answer emerges owing to cancellation of ambiguities
between the Sudakov exponent and the definition of the pole mass.
At the end the effect of the leading power correction, ${\rm
e}^{-(N-1)\bar{\Lambda}/M}$ in \eq{F_NP_result}, is to shift the perturbative spectrum
in~$x$ by the relative mass difference between the meson and the
quark. This result is important for phenomenology, as it opens up
the possibility to fix the leading non-perturbative correction
without using the data.

The DGE perturbative prediction for moments of the photon energy
spectrum in the $B\longrightarrow X_s\gamma$ decay is summarised
by~\eq{BXg_logs_N_2loop} with the matching coefficient
of~\eq{matching}. After taking the principal--value Borel integral
this result can be readily converted to $x$ space by an inverse
Mellin transform. It is important to note that having performed a
principal--value sum, the result is free of any Landau
singularity~(see \cite{CG}). This stands in sharp contrast with
the standard fixed--logarithmic--accuracy approximation, 
e.g.~\cite{Akhoury:1995fp,LR,LLR}, which
always presents such a singularity at $N\sim {\cal O}(M/\Lambda)$.
In practice, it is probably most convenient to include the parametrization of
non-perturbative corrections, \eq{F_NP_result}, directly in moment
space and then convert to $x$ space according to \eq{back_to_x} in
order to compare with data. 

Let us recall at this point that having 
concentrated on the specific problems of the endpoint region, we neglected 
other effects which are important for 
the phenomenology of radiative decays. One example is running--coupling effects, which 
have been shown  \cite{Ligeti:1999ea} to be important away from the endpoint
at order $\alpha_s^2$. Such corrections can be extracted 
from \eq{BXg_full} to any order. Other ${\cal O}(\alpha_s^2)$ corrections, which are 
harder to compute, may eventually be non-negligible.   
Obviously, one should also consider 
the contribution of operators other than the
magnetic one (\eq{O7}) on both the 
perturbative and non-perturbative levels. These involve 
qualitatively different effects. For example,
the operator $(\bar{c}_{L\alpha}\gamma^\mu b_{L\alpha})
(\bar{s}_{L\beta}\gamma_\mu c_{L\beta})$
involves a non-perturbative contribution associated with the 
production of an intermediate $J/\Psi$ state.
  
The analogous DGE perturbative prediction for the
moments of the charged--lepton energy spectrum in the
$\bar B\longrightarrow X_u l^-{\bar{\nu}}_l$ decay is summarised
by~\eq{SL_logs} and the matching coefficient of~\eq{C_sl}. As
previously discussed, non-perturbative corrections corresponding to the 
b-quark distribution in the B meson are the same as in the 
$\bar B\longrightarrow X_s\gamma$ decay. We note that other
differential distributions in the $\bar B\longrightarrow X_u
l^-{\bar{\nu}}_l$ decay, such as the distribution in the invariant
mass of the hadronic system near the endpoint, can be readily
computed from \eq{sl_logs_fact} by phase--space integration  
and exponentiation in the appropriate moment space.

On the theoretical side it is interesting to observe the universal
structure of the perturbative soft and jet functions. These two
functions were shown to be the same for the semi-leptonic
decay and the radiative $\bar{B}\longrightarrow X_s\gamma$ decay 
involving the magnetic operator; the difference between the two cases is restricted to the relations between the arguments of the functions and the scaling variables ({\em cf.} \eq{sl_logs_fact} and \eq{BXg_logs_fact}). 
Moreover, this perturbative universality extends beyond the application to
B decay. The soft function of \eq{Soft} represents radiation off
a heavy quark which is close to its mass shell and thus it also describes the
perturbative contribution to heavy--quark fragmentation~\cite{CG}.
The jet function describes radiation off an unresolved jet with a
constrained invariant mass and it is particularly important in the
analysis of deep inelastic structure function at large
Bjorken~$x$, where it is the only source of Sudakov
logs~\cite{Gardi:2001di,Gardi:2002bk,Gardi:2002xm}. It also
appears in single--particle inclusive cross sections in $e^+e^-$
annihilation for light \cite{Gardi:2001di} and heavy quarks
\cite{CG} as well as in event--shape
distributions~\cite{Gardi:2001ny,Gardi:2003iv}.


\acknowledgments

I would like to thank Gregory Korchemsky for suggesting this project to me, 
for sharing with me his insight on the problem and for many hours of discussion.
I also want to thank Volodya Braun and Bryan Webber for useful discussions. 
This work is supported by a Marie Curie individual
fellowship, contract number HPMF-CT-2002-02112.

\newpage

\appendix

\section{{\label{App_BXSG}} Extracting singular terms
in $\bar B\longrightarrow X_s\gamma$}

Here we summarize the formulae used to compute the singular terms
in the $x\longrightarrow 1$ limit from the full
single--dressed--gluon result of \eq{BXg_full}. In order to deal
with the Feynman--parameter integrals we use asymptotic expansions
of the form:
\begin{eqnarray}
\label{asym} 
&&\hspace*{-10pt}\int_0^1 d\alpha \frac{\alpha}{1-x\alpha}\nonumber
(1-\alpha)^{-u}=\frac{-1}{u(1-u)}\left(1+\frac{{1-x}}{1+u}+{\cal
O}\left(({1-x})^2\right)\right)
+\,\frac{\pi}{\sin\,\pi u}
\,(1-x)^{-u}x^{u-2},\nonumber \\
&&\hspace*{-10pt}\int_0^1 d\alpha \frac{\alpha^2}{(1-x\alpha)^2}
(1-\alpha)^{-u}=\frac{1}{u(1-u)(1+u)}\left(1+{\cal
O}(\left(1-x\right))\right)
\\ \nonumber && \hspace*{140pt}+\,\frac{\pi}
{\sin\,\pi u}\,(1-x)^{-u-1}x^{u-3}\left(u-2(1-x)\right).
\end{eqnarray}
The result is:
\begin{eqnarray}
\label{BXg_logs} \frac{1}{\Gamma_{\tot}}\frac{d\Gamma}{dx} &=&
(1+{\cal O}(\alpha_s)) \delta(1-x) +\frac{C_F}{2\beta_0}\int_0^{\infty}du
\left(\frac{\Lambda^2}{m^2}\right)^u\,{\rm e}^{cu}\,\frac{\sin\pi u}{\pi u} (1-x)^{-u}\nonumber\\
&& \bigg[-\frac{2\pi u}{\sin \pi
u}\,(1-x)^{-1-u}+\frac{2}{1-x}\left(-\frac{1}{u(1-u)}\,
+(1-x)^{-u}\frac{\pi}{\sin\pi u}\right)\nonumber\\
&& \qquad+\frac{1}{1-x}\,\frac{1}{(2-u)(1-u)} \bigg].
\end{eqnarray}
The three terms correspond to the diagrams I, II and III (in the
Feynman gauge), respectively. Organizing the result based on the
parametric dependence on $(1-x)$ we get \eq{BXg_logs_fact}.

\section{{\label{App_sl}} Renormalon calculation of the semi-leptonic decay spectrum}

\subsection{\label{sl_calc}A renormalon calculation}

Here we perform the single--dressed--gluon renormalon calculation
of the semi-leptonic decay spectrum with Borel--modified gluon
propagator. The three one--loop diagrams of figure~\ref{fig:diagrams}, contracted with the leptonic tensor, are denoted
$A_{\rm I}$, $A_{\rm II}$ and $A_{\rm III}$.
Computing the traces and performing the integral over the gluon
momentum using Feynman parameters we arrive at:
\begin{eqnarray}
\label{T}
A_{\rm I}&=&-m^6\,
\int_0^1da\,a^u(1-a)^{-u}\int_0^1d\lambda(1-\lambda)
\left(-\bar{\mu}^2\right)^{-u-2}
\sum_{n=0}^{3}k^{\rm I}_n(a,z,y,x,u)\lambda^n,\nonumber \\
A_{\rm II}&=&-m^4\, \int_0^1da\,a^u(1-a)^{-u}
\int_0^1d\lambda\left(-\bar{\mu}^2\right)^{-u-1}\times
\\
&&\hspace*{100pt}
\bigg[\frac{m^2}{l^2+i0}\sum_{n=0}^{2}\tilde{k}^{\rm
II}_n(a,z,x,u)\lambda^n +\sum_{n=0}^{2}k^{\rm II}_n(a,z,y,x,u)
\lambda^n\bigg],\nonumber \\\nonumber A_{\rm
III}&=&-m^2\,\frac{1}{u}\,
\int_0^1da\,a^u(1-a)^{-u}\left(\left.-\bar{\mu}^2
\right\vert_{\lambda=1}\right)^{-u}
\,\bigg[\frac{m^2}{l^2+i0}\tilde{k}^{\rm III}(a,z,x)+k^{\rm
III}(a,z,x)\bigg],
\end{eqnarray}
where we explicitly displayed the dependence on the Feynman
parameter $\lambda$, and $k_{n}$ and $\tilde{k}_n$ are polynomials
in the other dimensionless variables: the Feynman parameter $a$,
the external kinematic variables $y$, $z$ and $x$ and the Borel
parameter $u$. For the first diagram the coefficients are given by
\begin{eqnarray}
{k^{\rm I}_{0}} &=&  16\,(u + 1)\,(z - x)\,(y - x) - 16\,(u
 + 1)\,( - 2\,x + z)\,(y - x)\,(1 - a) \nonumber\\
&&\mbox{} + 8\,(2\,z\,(x + 1) - (z + x)\,(y - x)\,u - 2\,y\,z)\,(1
 - a)^{2} \nonumber\\
&&\mbox{} - 8\,( - 2\,z - 2\,x^{2} + 2\,y\,x + x\,(y - x)\,u)\,(1
 - a)^{3}
 \nonumber\\
{k^{\rm I}_{1}} &=& 8\,( x( - 1 + y - z)(y+z-x)
+ (y + 2\,u)\,z^{2} - (2\,y\,u + y^{2} + 2 - y)
\,z)(1 - a) \nonumber\\
&&\mbox{} + 8 ((2\,y - z - 1)\,x^{2} - (2\,z\,u - y + 2\,y^{2} + 2\,z)\,x + 2\,
z^{2}\,u \nonumber \\
&&\mbox{}+ (2\,y\,u + 2\,y - 2 + y^{2})\,z)(1 - a)^{2}  - 8\,( - z\,(x + y)\,u + y\,x^{2} + 2\,z\,x -
y^{2}\,x)\,(1 - a)^{3}
\nonumber\\
{k^{\rm I}_{2}} &=&8\,z\,( - ( - x^{2} + 2\,z + z\,x)\,u +
(3\,z - 2\,y + 3)\,x + y^{2} + ( - 1 - 2\,z)\,y)\,(1 - a)^{2} \nonumber\\
&&\mbox{} - 8\,z\,( - ( - 2\,z - x^{2})\,u + y\,(y - 2\,x))\,(1 - a
)^{3}
\nonumber \\
{k^{\rm I}_{3}} &=& 8\,(z^{2}\,x\,u + z^{2}\,(y - 2\,x))\,(1
 - a)^{3}
\end{eqnarray}
for the second by
\begin{eqnarray}
{\tilde{k}^{\rm II}_{0}} &=& 8\,a\,(z - x)\,(z^{2} + z - a\,z
 - z\,x - 1 + x) + {\displaystyle \frac {8\,(1 - a)^{2}\,( - x +
1 + z)\,(z - x)}{u}}
 \nonumber \\
{{\tilde{k}}^{\rm II}_{1}} &=&- 8\,( - 1 - z - a\,x + x + a\,
z + 2\,a)\,z\,(1 - a)\,(z - x) \nonumber \\
&&\mbox{}\hspace*{150pt}- {\displaystyle \frac {8\,(1 - a)^{2}\,(1 + z)\,( - x +
1 + z)\,(z - x)}{u}}
 \nonumber\\
{{\tilde{k}}^{\rm II}_{2}} &=&- 8\,(z - x)\,(1 - a)^{2}\,z\,(
 - x + 1 + z) + {\displaystyle \frac {8\,(z - x)\,(1 - a)^{2}\,z
\,( - x + 1 + z)}{u}}
   \nonumber\\
{{k}^{\rm II}_{0}} &=& - 8\,a\,z^{2} - 8\,a\,( - a + y - 2\,x
)\,z - 8\,a\,x - {\displaystyle \frac {8\,(z - x)\,(1 - a)^{2}}{u
}}
 \nonumber \\
{{k}^{\rm II}_{1}} &=&  8\,(1 - a)\,z\,( - z + a\,z + a\,y - 2\,a\,x + x)
\nonumber \\
&&\hspace*{150pt}- {\displaystyle \frac {8\,(1 - a)\,(z - x)\,( - z + a\,z
 + a - 1 - a\,x + a\,y)}{u}}  \nonumber \\
{{k}^{\rm II}_{2}} &=&  8\,(z - x)\,(1 - a)^{2}\,z -
{\displaystyle \frac {8\,(z - x)\,(1 - a)^{2}\,z}{u}}
\end{eqnarray}
and for the third by
\begin{eqnarray}
{\tilde{k}^{\rm III}} &=&  - 8\,a\,(z - x)\,(1 - x  + z)  \nonumber \\
{k^{\rm III}} &=& 8\,a\,(z - x)  .
\end{eqnarray}
The scale $-\bar{\mu}^2$ appears after the gluon momentum integration.
It is given by:
\begin{eqnarray}
\label{mu_bar}
-\bar{\mu}^2&=&\lambda^2(1-a)q^2-\lambda\left(q^2+m^2(1-ya)\right) +(1-a)m^2\nonumber \\
&=&-m^2(1-a)z\bigg[-\lambda^2+\lambda\frac{z+1-ya}{(1-a)z}-\frac1z\bigg]
\nonumber \\
&=&-m^2(1-a)z\bigg[(\lambda_2-\lambda)(\lambda-\lambda_1)\bigg]
\end{eqnarray}
For the diagram III it is evaluated at $\lambda=1$, so it simplifies a lot:
\[
\left.-\bar{\mu}^2\right\vert_{\lambda=1}=-l^2a.
\]
In \eq{mu_bar} we defined $\lambda_2$ as the larger of the two
(real) roots of the equation $-\bar{\mu}^2(\lambda)=0$. It turns
out that in the physical range $0<\lambda_1<1$ and $\lambda_2>1$,
so $-\bar{\mu}^2$ is negative only for $\lambda$ between
$\lambda_1$ and $1$. This is the real--emission contribution to
the imaginary part. Additional contributions come, as at the Born
level, from the $1/(l^2+i0)$ propagator terms.

Finally, let us explain how the $\lambda$ integrals can be
performed. When taking the imaginary part the integrals over
$\lambda$ range between $\lambda_1$ and $1$. They have the general
form:
\begin{eqnarray}
I_n(\omega)&=&\int_{\lambda_1}^1
d\lambda \,\lambda^n \,\left[(\lambda_2-\lambda)(\lambda-\lambda_1)\right]^\omega.
\end{eqnarray}
The result can be expressed in terms of hypergeometric functions. In particular,
\begin{eqnarray}
I_0(\omega)&=&\int_{\lambda_1}^1
d\lambda \left[(\lambda_2-\lambda)(\lambda-\lambda_1)\right]^\omega\nonumber \\
&=&\frac{1}{1+\omega}\left(\frac{a(1-y+z)}{(1-a)z}\right)^{\omega+1}\frac{(1-a)z}
{\sqrt{\Delta^2}}\times
\nonumber \\
&&\,\,\, F\left(\left[\frac12, \omega+1\right],[\omega+2],{4(1-y+z)a(1-a)z}/{\Delta^2}\right)
\end{eqnarray}
where $\Delta^2\equiv {(z+1-ya)^2-4(1-a)^2z}$.
We also note that
\begin{equation}
I_1(\omega)=-\frac12 \frac1{1+\omega}
\left(\frac{a(1-y+z)}{(1-a)z}\right)^{\omega+1}+
\frac{\lambda_1+\lambda_2}{2}I_0(\omega)
\end{equation}
and that any higher $n$ integral can be obtain using the following recursion relation:
\begin{eqnarray}
I_n(\omega)=-I_{n-2}(\omega+1)+(\lambda_1+\lambda_2)I_{n-1}(\omega)-\lambda_1
\lambda_2I_{n-2}(\omega).
\end{eqnarray}
To leading order in $(1-y+z)$ there is a significant
simplification:
\begin{equation}
I_n(\omega)\simeq
\frac{1}{1+\omega}\left(\frac{a(1-y+z)}{(1-a)z}\right)^{\omega+1}\frac{z}{1-z}+{\cal
O}(1-y+z), \label{I_n_approx}
\end{equation}
for {\em any} $n$. As we shall see below this approximation is
useful to extract the jet function while it is not valid for the
soft function, where higher--order terms in the expansion are
accompanied by higher singularities in $(1-a)$.

\subsection{\label{JETSOFT}Extracting singular terms}

Our purpose is to perform DGE in the triple differential rate in
semi-leptonic decays. To this end we need to extract all the terms
containing powers of~$\ln(1-y+z)$, to any logarithmic accuracy
(but to leading order in the large--$\beta_0$ limit), neglecting
terms that are suppressed by a power of $(1-y+z)$. It should be
emphasized that this does not imply we neglect powers of $(1-x)$: 
the dependence on $x$ remains at this stage exact.

In contrast with the radiative decay case, our calculation here is
guided by what we know about answer, namely that it contains two
scales from different kinematic origin: the jet mass scale $l^2$
which is proportional to $m^2(1-y+z)$ and the soft scale
$m(1-y+z)$; in the Borel representation the running coupling
appears as $(m^2/\Lambda^2)^u$, thus the former will generate
dependence on $(1-y+z)^{-u}$ while the latter dependence on
$(1-y+z)^{-2u}$.  Therefore, we would like to devise methods to
extract from the full result of section \ref{sl_calc} the singular
terms at~$y\longrightarrow 1+z$ and compute the two corresponding
Sudakov exponents separately, in section \ref{JET} the jet
function and in section \ref{SOFT} the soft one.

\subsubsection{\label{JET}Jet function}

We work in the Feynman gauge. Let us begin with diagram I in
figure~\ref{fig:diagrams}. We recall that in the radiative decay
this diagram contributes only to the soft function, i.e. it
involves the scale $m(1-y+z)$ but not $m^2(1-y+z)$. We will see that
this is so also here.

An exact calculation of diagram I yields (see \eq{T}):
\begin{equation}
A_{\rm I}=-m^6\,
\int_0^1da\,a^u(1-a)^{-u}\int_0^1d\lambda(1-\lambda)
\left(-\bar{\mu}^2\right)^{-u-2}
\sum_{n=0}^{3}k^{\rm I}_n(a,z,y,x,u)\lambda^n, \label{T_I}
\end{equation}
where $\bar{\mu}^2$ is the scale that appears when combining the
propagators by Feynman parametrization (\eq{mu_bar}):
\begin{eqnarray}
\label{mu_bar_text}
-\bar{\mu}^2&=&-m^2\bigg[-z(1-a)\lambda^2+(1+z-ya)\lambda-(1-a)\bigg],
\end{eqnarray}
and $k^{\rm I}_n$ are polynomials. The imaginary part
of $A_{\rm I}$ comes from the region where $\bar{\mu}^2>0$ so
$\lambda_1<\lambda<1$, where $\lambda_1$ is the smaller of the two
(real) solutions of $\bar{\mu}^2=0$. Since $\bar{\mu}^2$ is
quadratic in $\lambda$ the integral looks difficult. Expanding
$\bar{\mu}^2$ for $y\longrightarrow 1+z$ is not allowed, since
then $\lambda_1=1$ and the integral loses its support. Therefore
the $\lambda$ integral must be performed exactly. This can be done: 
the method is summarized at the end of the previous section.

Having performed the $\lambda$ integral, the leading dependence on
$l^2$ becomes explicit, through the factor $(1-y+z)^{-1-u}$.
Therefore, the jet function can be readily computed by
expanding\footnote{Such an expansion is not valid for the
calculation of soft terms, which are sensitive to the region
$a\longrightarrow 1$. We shall deal with these terms below.} in
powers of $(1-y+z)$ under the integral over $a$. To leading power in
$(1-y+z)$ one can use \eq{I_n_approx} so the integral
\[
\int_0^1d\lambda\lambda^n\left(-\bar{\mu}^2\right)^{-u-2}
\]
does not depend on $n$. One immediately realizes that owing to the
factor $1-\lambda$ in \eq{T_I} diagram I does not contribute to
the jet function at all.

Next, let us consider diagram II. Upon neglecting terms in the
numerator which are suppressed by a power of $(1-y+z)$, the second
line in \eq{T} reduces to\footnote{Note that we ignore here the
prescription for the light--quark propagator: when taking the
imaginary part we will be interested in the gluon emission cut, not 
in the pole.}:
\begin{equation}
A_{\rm II}\simeq \frac{-m^4}{(1-y+z)}\,
\int_0^1da\,a^u(1-a)^{-u}\int_0^1d\lambda\left(-\bar{\mu}^2\right)^{-u-1}
\sum_{n=0}^{2}\tilde{k}^{\rm II}_n(a,z,x,u)\lambda^n.
\label{T_II}
\end{equation}
Let us now compute the jet contribution to diagram II in the
method explained above. Using~\eq{I_n_approx} we have
\begin{eqnarray}
&&\int_0^1d\lambda\left(-\bar{\mu}^2\right)^{-u-1}
\sum_{n=0}^{2}\tilde{k}^{\rm II}_n(a,z,x,u)\lambda^n
\simeq\int_0^1d\lambda\left(-\bar{\mu}^2\right)^{-u-1}
\sum_{n=0}^{2}\tilde{k}^{\rm II}_n(a,z,x,u)\nonumber \\&&
\hspace*{30pt}
\simeq -8(x-z)(1+z-x) \left(-m^2\right)^{-1-u}a^{1-u}(1-a)^{-1}\frac{1}{u}\Big[(1-y+z)\Big]^{-u},
\end{eqnarray}
so the integral over $a$ is straightforward and the final result
is:
\begin{equation}
\left.A_{\rm II}\right\vert_{\rm jet}=
-8(x-z)(1+z-x)\left(-m^2\right)^{1-u}
\frac{1}{u^2(1-u)}\Big[(1-y+z)\Big]^{-1-u}.
\end{equation}

Diagram III is simple as it depends on just one external momentum,
$l$. The leading contribution in $(1-y+z)$ is:
\begin{eqnarray}
 A_{\rm III}&\simeq&{\displaystyle  {\,8(x-z)(1+z-x)\,}}\left(-m^2\right)^{1-u}
 \,\frac{1}{u}\,\Big[1-y+z\Big]^{-1-u}\,
\int_0^1 da\,a(1-a)^{-u}\nonumber \\&=&{8(x-z)(1+z-x)\,}\left(-m^2\right)^{1-u}
 \,\frac{1}{u(1-u)(2-u)}\,\Big[1-y+z\Big]^{-1-u}.
\end{eqnarray}
Clearly, this diagram contributes only to the jet function, not to
the soft one.

Finally, collecting the contributions of the different diagrams
and taking the imaginary part by replacing $(-1)^{1+u}$ by $\sin
\pi u$ we obtain:
\begin{eqnarray}
\label{Im_T_total_jet}
{\rm Im} \Big\{A_{\rm jet}\Big\}&=&{\rm Im}\Big\{2A_{\rm II}^{\rm jet}+A_{\rm III}^{\rm jet}\Big\}
\\ \nonumber
&=&-8\pi (x-z)(1+z-x)\left(m^2\right)^{1-u}\,\frac{\sin \pi u}{\pi
u}\left(\frac{2}{u}+\frac{1}{1-u}+\frac{1}{2-u}\right)\Big[1-y+z\Big]^{-1-u}.
\end{eqnarray}

\subsubsection{\label{SOFT}Soft function}

Consider now the singular contributions from diagram I on the soft
scale $m(1-y+z)$. These terms were discarded in the procedure
described above: the additional dependence on $(1-y+z)$ is
associated with the singular limit $a=1$. This suggest that the
soft terms can be computed by expanding \eq{T_I} near $a=1$: a
power of $(1-a)$ translates into a power of $(1-y+z)$. Let us
therefore replace the factor $a^u$ by~$1$ and then extend the
integration over $a$ in the lower limit to $-\infty$. This will
allow us to rescale the integration variable. We verified that
this manipulation does not change the imaginary part of the
answer.

Trading the integration over $a$ by a new variable $\xi$,
\begin{equation}
a=1-\frac{\xi(1+z-y)\lambda}{\lambda^2z-y\lambda+1},
\end{equation}
\eq{T_I} can be written as:
\begin{eqnarray}
\left.A_{\rm I}\right\vert_{\rm soft}&\simeq& 16(x-z)(1+z-x)\,(1+u)
\left(-m^2\right)^{1-u}(1+z-y)^{-1-2u}\,\times\nonumber \\
&&\int_0^1d\lambda
(1-\lambda)\lambda^{-1-2u}(\lambda^2z-y\lambda+1)^{u-1}
\int_0^{\infty}d\xi\,\xi^{-u}(1-\xi)^{-2-u} , \label{T_I_soft}
\end{eqnarray}
where we have changed the order of integration and approximated
the polynomial in the numerator by its leading power in the limit
$y\longrightarrow 1+z$ and thus also $a\longrightarrow 1$,
\[\sum_{n=0}^{3}k^{\rm I}_n(a,z,y,x,u)\lambda^n\,\,\longrightarrow\,\,
-16(x-z)(1+z-x)\,(1+u).\]
This way we managed to factorize the difficult double integral
into a product of two simple integrals. Computing the diagram in
the soft (Eikonal) approximation one immediately obtains
\eq{T_I_soft}.

Coming to evaluate \eq{T_I_soft} we note that only the region
$0<\xi<1$ contributes to the imaginary part, and the $\xi$ integral
becomes
\[
\int_0^{1}d\xi\,\xi^{-u}(1-\xi)^{-2-u}=-2^{1+2u}\sqrt{\pi}\,u\,\Gamma(-1-u)/\Gamma
\left(\frac12-u\right),
\]
while the $\lambda$ integral is
\[
\int_0^1d\lambda
(1-\lambda)\lambda^{-1-2u}(\lambda^2z-y\lambda+1)^{u-1}=-2^{-1-2u}(1-z)^{2u}
\Gamma(u)
\Gamma\left(\frac12-u\right)/\sqrt{\pi}+{\cal O}(1-y+z).
\]
Thus, the final answer for the soft contribution to $A_I$ is:
\begin{eqnarray}
\left.A_{\rm I}\right\vert_{\rm
soft}&\simeq& 16 (x-z)(1+z-x)\,(1-z)^{2u}\left(-m^2\right)^{1-u}
\frac{\pi}{\sin\pi u}\,\Big[1-y+z\Big]^{-1-2u}.
\label{T_I_soft_final}
\end{eqnarray}

Consider now the soft contribution of diagram II, \eq{T_II}. To
leading power in $(1-y+z)$, and thus also in $(1-a)$, the numerator
reduces to
\[
\sum_{n=0}^{2}\tilde{k}^{\rm
II}_n(a,z,x,u)\lambda^n = 8(x-z)(1+z-x)\,(1-z).
\]
Repeating the procedure explained above we get
\begin{eqnarray}
\hspace*{-20pt}
\left.A_{\rm II}\right\vert_{\rm soft}&=&
-8(x-z)(1+z-x)\,(1-z)\left(-m^2\right)^{1-u}
\Big[1-y+z\Big]^{-1-2u}\,\times\nonumber \\
&&\hspace*{30pt}\int_0^1d\lambda\lambda^{-2u}\left(\lambda^2z-\lambda
y+1\right)^{u-1}
\int_0^{\infty}d\xi\,\xi^{-u}(1-\xi)^{-u-1}\nonumber \\
&=&-8(x-z)(1+z-x)\,(1-z)^{2u}\left(-m^2\right)^{1-u}\,\frac{\pi}{\sin
\pi u}\,\frac{1}{u}\Big[1-y+z\Big]^{-1-2u}.
\end{eqnarray}
As already mentioned, diagram III does not contribute to the soft
function.

Finally, collecting the contributions of the different diagrams
and taking the imaginary part (replacing $(-1)^{1+u}$ by $\sin \pi
u$) we get:
\begin{equation}
{\rm Im} \Big\{A_{\rm soft}\Big\}={\rm Im}\Big\{A_{\rm I}^{\rm
soft}+2A_{\rm II}^{\rm
soft}\Big\}=16\pi (x-z)(1+z-x)(1-z)^{2u}
\left(m^2\right)^{1-u}\frac{1-u}{u}\Big[1-y+z\Big]^{-1-2u}.
\label{Im_T_total_soft}
\end{equation}

\end{document}